\documentclass[11pt]{article}
\usepackage{amsmath,amssymb,color,epsf}
\usepackage{amssymb,slashed,latexsym,amsmath,multirow,color}
\usepackage{graphics}
\makeatletter
\@addtoreset{equation}{section}
\makeatother

\usepackage{amsfonts}

\newcommand{\hoch}[1]{$\, ^{#1}$}

\usepackage{float}

\usepackage{soul}

\usepackage{amsfonts}

\usepackage[font=footnotesize,labelsep=newline,labelfont=sc,justification=centering,position=top]{caption}

\usepackage{float}

\newcommand{\be}{\begin{equation}}
\newcommand{\ee}{\end{equation}}
\newcommand{\bea}{\setlength\arraycolsep{2pt} \begin{eqnarray}}
\newcommand{\eea}{\end{eqnarray}}
\newcommand{\nn}{\nonumber}

\usepackage{soul}
\usepackage{yfonts}
\usepackage[font=footnotesize,labelsep=newline,labelfont=sc,justification=centering,position=top]{caption}

\usepackage{hyperref}

\newcommand{\ft}[2]{{\textstyle\frac{#1}{#2}}}

\newsavebox{\uuunit}
\sbox{\uuunit}
    {\setlength{\unitlength}{0.825em}
     \begin{picture}(0.6,0.7)
        \thinlines
        \put(0,0){\line(1,0){0.5}}
        \put(0.15,0){\line(0,1){0.7}}
        \put(0.35,0){\line(0,1){0.8}}
       \multiput(0.3,0.8)(-0.04,-0.02){12}{\rule{0.5pt}{0.5pt}}
     \end {picture}}

\def\be{\begin{equation}}
\def\ee{\end{equation}}
\def\ba{\begin{array}}
\def\ea{\end{array}}
\def\bea{\begin{eqnarray}}
\def\eea{\end{eqnarray}}
\def\bd{\begin{displaymath}}
\def\ed{\end{displaymath}}

\def\nn{\nonumber}

\def\g{\gamma}

\def\d{\delta}

\def\e{\epsilon}

\def\vf{\varphi}

\def\p{\psi}

\def\l{\lambda}
\def\L{\Lambda}
\def\m{\mu}
\def\n{\nu}
\def\r{\rho}
\def\s{\sigma}

\def\t{\tau}

\def\o{\omega}
\def\O{\Omega}

\def\nn{\nonumber}

\def\cD{\mathcal{D}}
\def\cN{\mathcal{N}}
\def\cA{\mathcal{A}}
\def\cL{\mathcal{L}}
\def\cF{\mathcal{F}}

\def\cR{\mathcal{R}}

\begin{document}

\begin{flushright}
\hfill{ \
\ \ \ \ UG-72-2014 \ \ \ \ }
\end{flushright}
\vskip 1.2cm

\begin{center}
{\Large \bf  3D Born-Infeld Gravity and Supersymmetry}
\end{center}
\vspace{25pt}
\begin{center}
{\Large {\bf }}

\vspace{10pt}

{\Large Eric Bergshoeff\hoch{} and Mehmet Ozkan\hoch{}
}
\vskip .2truecm
{{\tt e.a.bergshoeff@rug.nl, m.ozkan@rug.nl}}
\vskip .2truecm
\vspace{10pt}

\hoch{} {\it Institute for Particle Physics and Gravity, University of Groningen,\\
Nijenborgh 4, 9747 AG Groningen, The Netherlands}

\vspace{10pt}

\underline{ABSTRACT}

\end{center}

We construct the most general parity-even higher-derivative $\cN=1$ off-shell supergravity model in three dimensions with a maximum of six derivatives. Excluding terms quadratic in the curvature tensor with two explicit derivatives and requiring the absence of ghosts in a linearized approximation around an $\text{AdS}_3$ background, we find that there is a unique supersymmetric invariant which we call supersymmetric  `cubic extended' New Massive Gravity.
The  purely gravitational part of this invariant is in agreement with an earlier analysis based upon the holographic c-theorem and coincides with an expansion of Born-Infeld gravity to the required order.

Our results lead us to propose an expression for the bosonic part of off-shell  $\cN=1$ Born-Infeld supergravity in three dimensions that is free of ghosts. We show that different truncations of a perturbative expansion of this expression gives rise to the bosonic part of (i) Einstein supergravity; (ii) supersymmetric New Massive Gravity and (iii) supersymmetric `cubic extended' New Massive Gravity.

\vspace{15pt}

\thispagestyle{empty}

\vspace{15pt}

 \vfill

\thispagestyle{empty}
\voffset=-40pt

\newpage

\tableofcontents


\newpage


\section{Introduction}

Higher-derivative supergravity Lagrangians have received considerable attention in recent years. There are several reasons for this. First of all, such models have been used to calculate corrections to the BPS black hole entropy in a given setting \cite{Lopes Cardoso:1998wt}. More generally, such higher-derivative models provide an interesting test of the range of validity of the AdS/CFT correspondence \cite{Maldacena:1997re}. Higher-derivative gravity models have also occurred in  discussions of the holographic c-theorem \cite{Myers:2010tj}. More recently, a specific four-derivative gravity model in
three dimensions, called New Massive Gravity (NMG) \cite {Bergshoeff:2009hq}, has been introduced as the parity-even version of Topologically Massive Gravity (TMG) \cite{Deser:1981wh}.

An example of a 3D higher-derivative gravity theory with an infinite number of higher-derivative terms
is the so-called Born-Infeld (BI) gravity theory \cite{Gullu:2010pc} which is a particular example of the
Born-Infeld-Einstein theories studied in \cite{Deser:1998rj}.
It was observed in \cite{Gullu:2010pc} that upon making a truncation, that keeps terms with at most four derivatives in a perturbative expansion
of this BI gravity theory, one ends up with the ghost-free NMG model \cite{Bergshoeff:2009hq}. Furthermore,
by making a different truncation that retains  terms with at most six derivatives one ends up with
what we will call  the `cubic extended'  NMG model \cite{Gullu:2010pc,Sinha:2010ai, Paulos:2010ke}.

The supersymmetric extension of NMG  was given in \cite{Andringa:2009yc,Bergshoeff:2010mf}. The construction of such an invariant turns out to be more subtle than anticipated. It is not difficult to construct the $\cN=1$ supersymmetric extension
of the Einstein-Hilbert term  with a cosmological constant and of  the Ricci tensor  and Ricci scalar squared terms\,\footnote{Remember that in three dimensions the Riemann tensor is is proportional to the Ricci tensor.}
that make up the NMG invariant. When setting the cosmological constant to zero and linearizing the model around the maximally supersymmetric Minkowski$_{3}$ background, one finds that the model is ghost-free as expected. However, a complication arises when taking the cosmological constant into account. The Ricci scalar squared and the Ricci tensor squared invariants include couplings between the gravitational field and the auxiliary scalar $S$ that is part of the $\cN=1$ supergravity multiplet in such a way that ghosts are re-introduced in the linearized approximation around $\text{AdS}_3$.

To be specific, assigning the gravitational field a mass dimension zero and the auxiliary scalar $S$ a mass dimension one, the bosonic part of the  most general $\cN =1$ supersymmetric invariant of the form $1/m^2\, \times$ [terms of mass dimension four] is given by \cite{Andringa:2009yc,Bergshoeff:2010mf}
\bea\label{mass4}
e^{-1} \cL_{\text{SQuad}}
&=& \frac{1}{m^2} \Big[a_1 R_{\m\n} R^{\m\n} + a_{2} R^2 - ( 6 a_1 + 16 a_2  ) \partial_\m S \partial^\m S \nn\\
&& + \Big( 4 a_1 + 12 a_2 + \frac{3}{10} a_3 \Big) R S^2 + ( 12 a_1 + 36 a_2 + a_3 ) S^4 \Big]\,,
\eea
where $m$ is a mass parameter and $a_1,a_2,a_3$ are dimensionless parameters representing three different supersymmetric invariants. To obtain the complete model, the above invariant has to be added to the supersymmetric completion of the Einstein-Hilbert term with a cosmological constant.
We will call the $RS^2$ term that contains both the curvature tensor and the auxiliary scalar `off-diagonal'.

When four-derivative terms are added to the Poincar\'e supergravity, the auxiliary scalar $S$ becomes dynamical, and the dynamical scalar forms a multiplet with the gravitational scalar ghost, and the gamma trace of the gravitino field strength $\g \cdot \cR$ where $\cR^\m = \e^{\m\n\r} \nabla_\n \p_\r$. Therefore, the kinetic term for the auxiliary scalar $S$ in \eqref{mass4} is unacceptable since it would introduce an extra ghost degree of freedom. The coefficient in front of this kinetic term vanishes precisely for the NMG combination $a_2 = - \ft38 a_1$, in which case the $(\ft12, 0, 0)$ ghost multiplet becomes infinitely heavy and decouples from the spectrum. That still leaves us with the off-diagonal $RS^2$ term. It is easily seen that,
given the complete model, upon linearization  around an $\text{AdS}_3$ background with a constant value of the auxiliary scalar\,\footnote{We use here that such configurations are solutions of the complete model.}
the presence of the $RS^2$ term  leads, after elimination of the auxiliary scalar perturbation, to extra terms quadratic in the curvature tensor thereby upsetting the
ghost-free NMG combination. Therefore, to cancel the unwanted $RS^2$ term, we need to take $a_3 = \ft{5}{3}a_1$.
Taking $a_1=1$, we thus end up with the ghost-free combination
\bea
e^{-1} \cL_{\text{ghost-free}}
&=& \frac{1}{m^2} \Big[ \big( R_{\m\n} R^{\m\n} -\frac{3}{8}  R^2\big ) + \frac{1}{6}S^4\Big] \,,
\eea
that is part of the supersymmetric NMG model \cite{Andringa:2009yc,Bergshoeff:2010mf}.

In this paper we wish to push the above analysis one level further and consider the supersymmetric completion of the most general terms of the form $1/m^4\, \times$ [terms of mass dimension six].  We do not consider terms of the form
$R\Box R$ or $R_{\m\n}\Box R^{\m\n}$ since such terms lead to models with {\sl two} massive
gravitons one of which is a ghost \cite{Nutma:2012ss}.
It turns out that one can write down
15 terms of mass dimension 6 that involve the metric tensor and/or the auxiliary scalar. Considering only purely gravitational
terms that are cubic in the curvature tensor,
we find that a supersymmetric completion involves only 11 of these  terms\,\footnote{We do not consider the
terms $(\Box S)^2$ and $S\epsilon_{\m\n\rho}R^\mu{}_\sigma \nabla^\rho R^{\nu\sigma}$ since these terms only occur as part of a supersymmetric $R\Box R$ or $R_{\m\n}\Box R^{\m\n}$ invariant which we exclude.}.
Three of them are purely gravitational, six are off-diagonal and
two are purely auxiliary:
\bea
&&R_{\m\n} R^{\n\r} R_\r{}^\m\,,\,  R^3\,,\,  R R_{\m\n} R^{\m\n}:\hskip 1.1truecm \text{purely gravitational}\ (3)\,,\nonumber \\[.2truecm]
&&R^2 S^2\,,\, R S^4\,,\, S^2 R_{\m\n} R^{\m\n}\,,\\[.2truecm]
 &&R_{\m\n} \partial^\m S \partial^\n S\,,\, R \partial_\m S \partial^\m S\,,\, R S \Box S: \hskip .4truecm \text{off-diagonal} \ (6)\,,\nonumber\\[.2truecm]
&&S^2 \partial_\m S \partial^\m S\,,\, S^6:\hskip 3.3truecm \text{purely auxiliary}\ (2)\,.\nonumber
\eea
We find that the supersymmetric completion of these terms contains  8 parameters
corresponding to the 8 supersymmetric invariants that we will construct in this work.
One of the terms in the above expression ($S^2\partial_\mu S\partial^\mu S$) is un-acceptable since in a linearization it gives rise to a dynamical $S$ in the spectrum, which, once again forms a ghost multiplet with the gravitational scalar ghost and the gamma trace of the gravitino field strength. Furthermore, all of the 6 off-diagonal terms are un-acceptable
since each of them would upset, after linearization around an $\text{AdS}_3$ background, the ghost-free NMG combination.
This imposes a total of 7 restrictions on our 8 parameters thereby leading to a unique solution which is given in eq.~\eqref{SuperCNMG}.  This equation contains all terms of mass dimension 6 of what we will call the supersymmetric `cubic extended'  NMG model.

We find that the purely gravitational terms of the supersymmetric `cubic extended'  NMG model precisely coincides with
a specific truncation of BI gravity that  keeps terms with at most 6 derivatives.
 Given our results for the supersymmetric completion of this specific truncation of the BI gravity model
we are led to propose an expression for the bosonic part of the  $\cN=1$ BI supergravity theory, see eq.~\eqref{SDBI0}.
Upon making different truncations this $\cN=1$ BI supergravity model reproduces the bosonic sector of
(i) Einstein supergravity; (ii) supersymmetric New Massive Gravity and (iii) supersymmetric `cubic extended' New Massive Gravity. We will comment about some further features of this proposed BI supergravity model.

The organisation of this paper is as follows. As a warming up exercise we will briefly review in section \ref{s2}
how to construct Poincar\'e supergravity with a cosmological constant using the superconformal tensor calculus.
In section \ref{s3} we will consider the supersymmetric completion of the most general terms of mass dimension 4
thereby re-producing the supersymmetric NMG model \cite{Andringa:2009yc,Bergshoeff:2010mf}. The purpose of this section is to
prepare the reader for  section \ref{s4} where we will push the analysis one level further and consider the supersymmetric extension of the most general terms of mass dimension 6.  The results obtained in this section will enable us to construct in section \ref{ss:5} the supersymmetric completion of cubic extended NMG.
In a separate subsection,
based on our experience with the supersymmetric completion of the NMG and the cubic extended NMG models, we propose an expression for the bosonic part of $\cN =1$ Born-Infeld supergravity. We will comment about our results in the Conclusions.


\section{Poincar\'e Supergravity with a Cosmological Constant}\label{s2}

The purpose of this section is to familiarize the reader with the conformal tensor calculus and construct
Poincar\'e supergravity with a cosmological constant.
 In the first subsection, we introduce the $3D,~\cN=1$ Weyl multiplet and the basic matter multiplets we will need
 in this paper,
 i.e.~the scalar and Yang-Mills multiplet.
 In the second subsection, using these basic ingredients, we construct the Poincar\'e supergravity theory with a cosmological constant.
For an early reference on $3D$
supergravity, see \cite{Gates:1983nr}.
 Below, we will make extensive use of the  superconformal  $D=3,~\cN=1$ superconformal tensor calculus constructed in \cite{vanNieuwenhuizen:1985cx, Uematsu:1984zy, Uematsu:1986de}.

\subsection{Weyl and Matter Multiplets}

Our starting point is the $3D,~\cN=1$ Weyl multiplet, with $2+2$ off-shell degrees of freedom, whose independent gauge fields are the Dreibein $e_\mu{}^a$,
the gravitino $\psi_\mu$ and the dilatation gauge field $b_\mu$\,\footnote{$a=0,1$ is a Lorentz index and $\mu=0,1$ is a curved index. We use two-component Majorana spinors.}. These gauge fields transform under dilatations $D$, special conformal transformations $K$, $Q$-supersymmetry  and $S$-supersymmetry,
with parameters $\Lambda_D, \Lambda_K, \epsilon$ and $\eta$, respectively, as follows
\cite{Uematsu:1984zy}:
\bea
\d e_\m{}^a &=& \ft12 \bar\e \g^a \p_\m - \Lambda_D e_\mu{}^a\nn\,,\\
\d \p_\m &=& \cD_\m (\o) \e + \g_\m \eta -\ft12 \Lambda_D\psi_\mu \nn\,,\\
\d b_\m &=& - \ft12 \bar\e \phi_\m  + \ft12 \bar\eta \p_\m + \partial_\mu \Lambda_D + 2\Lambda_{K\mu}\,,
\label{WeylTransform}
\eea
where
\bea
\cD_\m (\o) \e = \Big( \partial_\m + \ft12 b_\m + \ft14 \o_\m{}^{ab} \g_{ab} \Big) \e
\eea
is covariant with respect to dilatations and Lorentz transformations.
The expressions for the spin-connection field $\omega_\mu{}^{ab}$, the special conformal  gauge field $f_\mu{}^a$ and the S-supersymmetry gauge field $\phi_\mu$
in terms of the independent gauge fields $e_\m{}^a, \p_\m$ and $b_\m$ are given by
\bea
\o_\m{}^{ab} &=& 2 e^{\n[a} \partial_{[\m} e_{\n]}{}^{b]} - e^{\n[a} e^{b]\s} e_{\m c} \partial_\n e_\s{}^c + 2 e_\m{}^{[a} b^{b]} + \ft12 \bar\p_{\m} \g^{[a} \p^{b]}  + \ft14 \bar\p^a \g_\m \p^b \,,\nn\\
\phi_\m &=& - \g^a \widehat{R}'_{\m a}(Q) + \ft14 \g_\m \g^{ab} \widehat{R}'_{ab} (Q) \,,\nn\\
f_\m{}^a &=&  - \ft12 \widehat{R}'_\m{}^a(M)  + \ft18 e_\m{}^a \widehat{R}'(M) \,,
\label{gaugefield}
\eea
where $\widehat{R}_{\m}{}^a(M) =  \widehat{R}_{\m\n}{}^{ab}(M) e^\n{}_b$ and $\widehat{R}(M) = e_a{}^\mu \widehat {R}_\mu{}^a(M)$.
Here  $\widehat{R}_{\m\n} (Q)$ and $\widehat{R}_{\m\n}{}^{ab}(M)$ are the super-covariant
curvatures of  $Q$-supersymmetry and Lorentz transformations $M$, respectively.
The expressions for these curvatures are given by
\bea
\widehat{R}_{\m\n}(Q) &=& 2 \partial_{[\m} \p_{\n]} + \ft12 \o_{[\m}{}^{ab} \g_{ab} \p_{\n]} + b_{[\m} \p_{\n]} - 2 \g_{[\m} \phi_{\n]} \,,\nn\\
\widehat{R}_{\m\n}{}^{ab}(M) &=& 2 \partial_{[\m} \o_{\n]}{}^{ab} + 2 \o_{[\m}{}^{ac} \o_{\n]c}{}^b + 8 f_{[\m}{}^{[a} e_{\n]}{}^{b]} - \bar\phi_{[\m} \g^{ab} \p_{\n]} \,.
\eea
The prime in $\widehat{R}'(M)$ indicates that we have omitted the $f_\m{}^a$ term in the expression for the full curvature $\widehat{R}(M)$. Similarly, the
prime in  $\widehat{R}'(Q)$ indicates that we have omitted the $\phi_\mu $ term in the expression for the full curvature $\widehat{R}(Q)$.
In table \ref{table1}, see below, we have collected a few  basic properties of the Weyl multiplet.

We next introduce the basic superconformal matter multiplets. First, we consider
the scalar multiplet with $2+2$ off-shell degrees of freedom. This multiplet consists of a physical scalar $\cA$, a Majorana fermion $\chi$ and an auxiliary scalar $\cF$. We assign the scalar $\cA$ an arbitrary dilatation weight $\o_\cA$. The weights of the other fields, together with some other properties of the scalar multiplet,
can be found in table \ref{table1}. The $Q$ and $S$ transformations of the scalar multiplet component fields are given by \cite{Uematsu:1984zy}
\bea
\d \cA &=& \ft14 \bar\e \chi \,,\nn\\
\d \chi &=& \slashed{\cD}\cA \e - \ft14 F \e - 2 \cA \o_\cA  \eta \nn\,,\\
\d \cF &=& - \bar\e\,\slashed{\cD} \chi - 2 (\o_\cA - \ft12) \bar\eta \chi \,,
\label{ScalarTransformation}
\eea
where the supercovariant derivatives of $\cA$ and $\chi$ are given by
\bea
\cD_\m \cA &=& ( \partial_\m - \o_\cA b_\m ) \cA - \ft14 \bar\p_\m \chi \,,\nn\\
\cD_\m \chi &=& \Big(\partial_\m - (\o_\cA + \ft12) b_\m + \ft14 \o_\m{}^{ab} \g_{ab} \Big) \chi - \slashed{\cD} \cA \p_\m + \ft14 F \p_\m + 2 \cA \o_\cA \phi_\m \,.
\label{ConformalDerivative}
\eea
In the rest of this paper we will need four special scalar multiplets with specific choices of the weight $\o_\cA$. It is convenient to give  names to the components of these four multiplets, see table \ref{special}.

{\small
\begin{table}[h]
\begin{center}
\begin{tabular}{|c|c|c|c|c|}
\hline
multiplet&field& type&off-shell&$D$\\[.1truecm]
\hline\rule[-1mm]{0mm}{6mm}
Weyl&$e_\mu{}^a$&dreibein&2&1\\[.1truecm]
&$\psi_\mu$&gravitino&2&$\ft12$\\[.1truecm]
\hline
Scalar&$\cA$&scalar&1&$\o_\cA$ \\[.1truecm]
&$\chi$&spinor&2&$\o_\cA +\ft12$\\[.1truecm]
&$\cF$&auxiliary&1&$\o_\cA+1$ \\[.1truecm]
\hline\rule[-1mm]{0mm}{6mm}
Yang-Mills&$A_\mu^I$&gauge field&2&0\\[.1truecm]
&$\vf^I$&gaugino&2&$\ft32$\\[.1truecm]
\hline
\end{tabular}
\end{center}
\caption{\footnotesize This table summarizes some properties of the basic multiplets of the $3D$ superconformal tensor calculus. The fourth column indicates the number of off-shell degrees of freedom represented by the field. The last column specifies the dilatation weight. For the Yang-Mills multiplet we have
only counted the off-shell degrees of freedom for one value of the Lie algebra index $I$.}\label{table1}
 \end{table}
 }

We finally consider the
superconformal Yang-Mills multiplet. This multiplet consists  of a vector field $A_\m^I$, with zero Weyl weight, and a Majorana spinor $\vf^I$, where $I$ is a Lie algebra index, see table \ref{table1}. This multiplet has $2+2$ off-shell degrees of freedom per value of the Lie algebra index. The supersymmetry transformations of the component fields are given by
\bea
\d A_\m^I &=& -\bar\e \g_\m \vf^I \,,\nn\\
\d \vf^I &=& \ft18 \g^{\m\n} \widehat{F}_{\m\n}^I \e \,,
\eea
where the supercovariant field strength is defined as
\bea
\widehat{F}_{\m\n}^I  &=& 2 \partial_{[\m} A_{\n]}^I + g f_{JL}{}^I A_\m^J A_\n^L + 2 \bar\p_{[\m} \g_{\n]} \vf^I \,.
\eea

{\small
\begin{table}[h]
\begin{center}
\begin{tabular}{|c|c|c|}
\hline
name& $\o_\cA$& components\\[.1truecm]
\hline\rule[-1mm]{0mm}{6mm}
Auxiliary Scalar Multiplet&2&$(Z,\O,F)$ \\[.1truecm]
Curvature Scalar Multiplet&$\ft32$&$(\xi,\vf,M)$\\[.1truecm]
Compensating Scalar Multiplet &$\ft12$&$(\phi,\lambda,S)$\\[.1truecm]
Neutral Scalar Multiplet&0&$(\sigma,\psi,N)$ \\[.1truecm]
\hline
\end{tabular}
\end{center}
\caption{\label{special}\footnotesize  This table indicates the nomenclature for the four scalar multiplets that will play a special role in the superconformal tensor calculus we use in this paper.}
\end{table}
 }

\subsection{Poincar\'e Supergravity with a Cosmological Constant}

In this subsection we construct a Poincar\'e supergravity theory, together with a
cosmological constant, using the superconformal ingredients given in the previous subsection.
Our starting point is the action for a general scalar multiplet coupled to conformal supergravity.
To cancel the weight --3 of the $e =\text{det}\,e_\mu{}^a$ factor in the leading term, we need to consider an auxiliary scalar multiplet since its highest component $F$ has weight 3. The Lagrangian corresponding to such an action is given by \cite{Uematsu:1984zy}\,\footnote{This is the full Lagrangian including the fermionic terms. From now on, to avoid the cluttering of complicated expressions, we adopt the convention that we only give the bosonic part of a supersymmetric Lagrangian
unless we explicitly state  that we include the fermions as we will do in a few cases. If needed, all fermionic terms can easily be restored by the superconformal tensor calculus.}
\bea
e^{-1} \cL_{\text{Aux}} &=& F -  \bar\p_\m \g^\m \O - Z \bar\p_\m \g^{\m\n} \p_\n \,.
\label{Weight2Action}
\eea

Before proceeding to the construction of the Poincar\'e supergravity action, we first discuss the multiplication rule for scalar multiplets and the construction of a kinetic multiplet for any given scalar multiplet. The multiplication rule states that  two scalar multiplets can be multiplied together to form a third scalar multiplet. In component formalism, if we consider two scalar multiplets  $(\cA_i, \chi_i, \cF_i )$, $i = 1,2$, with the conformal weight of the lowest component given by $\o_{\cA_i}$ a third scalar multiplet with components $\big(\cA_3\,,\chi_3\,,\cF_3\big)$ and weight $\o_{\cA_3} = \o_{\cA_1} + \o_{\cA_2}$ can be obtained by the following multiplication rules:
\bea
\cA_3 &=& \cA_1 \cA_2 \,,\nn\\
\chi_3 &=& \cA_1 \chi_2 + \cA_2 \chi_1 \,,\nn\\
\cF_3 &=& \cA_1 \cF_2 + \cA_2 \cF_1 + \bar\chi_1 \chi_2 \,.
\label{multip}
\eea
These rules can be inverted as follows:
\bea
\cA_1 &=& \cA_3 \cA_2^{-1} \,,\nn\\
\chi_1 &=& \cA_{2}^{-1} \chi_3 - \cA_3 \cA_2^{-2} \chi_2 \,,\nn\\
\cF_1 &=& \cA_2^{-1} \cF_3 - \cA_3 \cA_2^{-2} \cF_2 - \cA_2^{-2}  \bar\chi_2 \chi_3 + \cA_3 \cA_2^{-3} \bar\chi_2 \chi_2 \,.
\label{decomp}
\eea
Next, we discuss the idea of a kinetic multiplet of a given scalar multiplet. A kinetic multiplet is formed from the composite expressions of a scalar multiplet in which the dynamical scalar $\cA$ and fermion $\chi$ have a kinetic term.  In terms of components, given a scalar multiplet, indicated with the subscript $s$, a kinetic multiplet, denoted
 by the subscript $k$, is obtained by the following rules:
\bea
\cA_k &=& \o_{\cA_s} \cF_s - \ft12 \Big( \o_{\cA_s} - \ft12 \Big) \cA_s^{-1} \bar\chi_s \chi_s \,,\nn\\
\chi_k &=& - 4  \o_{\cA_s} \slashed{\cD} \chi_s + 4 \Big( \o_{\cA_s} - \ft12 \Big) \cA_s^{-1} \slashed{\cD} \cA_s \chi_s + \Big( \o_{\cA_s} - \ft12 \Big) \cA_s^{-1} \cF_s \chi_s \nn\\
&& + \ft12 \Big( \o_{\cA_s} - \ft12 \Big) \cA_s^{-2} \chi_s \bar\chi_s \chi_s \,,\nn\\
\cF_k &=& 16  \o_{\cA_s} \Box^c \cA_s - 16 \Big( \o_{\cA_s} - \ft12 \Big) \cA_s^{-1} \cD_\m \cA_s \cD^\m \cA_s + \Big( \o_{\cA_s} - \ft12 \Big) \cA_s^{-1} \cF_s^2 \nn\\
&& -4 \Big( \o_{\cA_s} - \ft12 \Big) \cA_s^{-1} \bar\chi_s \slashed{\cD} \chi_s  -\ft12 \Big( \o_{\cA_s} - \ft12 \Big) \cA_s^{-2} \cF_s \bar\chi_s \chi_s   \nn\\
&&- \ft12 \Big( \o_{\cA_s} - \ft12 \Big) \cA_s^{-3} \bar\chi_s \chi_s \bar\chi_s \chi_s \,.
\label{kinet}
\eea
The superconformal d'Alambertian occurring in the expression for $\cF_k$ is given by
\bea
\Box^c \cA_s &=& (\partial^a - (\o_s + 1) b^a + \o_b{}^{ba} ) \cD_a \cA_s + 2 \o_s f_a^a \cA_s - \ft14 \bar\p_a \cD^a \chi_s - \ft14 \bar\phi_a \g^a \chi_s \,.
\eea
For future reference, we notice that
applying   the kinetic multiplet formula (\ref{kinet}) we obtain  the following relation between a compensating multiplet   $\big(\phi\,,\lambda\,,S\big)$ and  a curvature multiplet
 $\big(\xi\,,\varphi\,,M\big)$:
\bea
\xi = S \,,\hskip 1truecm
\vf  = - 4 \slashed{\cD} \l \,,\hskip 1truecm
M = 16 \Box^c \phi \,, \label{mapI}
\label{CurvtoComp}
\eea
where we have rescaled the composite formulae (\ref{kinet}) with a factor of 2. Notice that the difference between the conformal weights of the lowest components of the scalar and corresponding kinetic  multiplets is $\o_k - \o_s = 1$. Using the kinetic multiplet and the (inverse) multiplication rules, it is not difficult to obtain the Lagrangian for a scalar multiplet with conformal weight $\o$ coupled to a compensating multiplet with
components $\big(\phi\,,\lambda\,, S\big)$, see table \ref{special}. One first multiplies the scalar multiplet with its kinetic multiplet and, next, with appropriate powers of the compensating multiplet. One thus obtains the following result:
\bea
e^{-1} \cL_\o &=& 16 \o \phi^{2 - 4 \o} \cA_\o \Box^c \cA_\o - 16 \Big( \o - \ft12 \Big) \phi^{2 - 4\o} \cD_\m \cA_\o \cD^\m \cA_\o \nn\\
&& + \phi^{2 - 4\o} \Big( 2\o - \ft12 \Big) \cF_\o^2 - \o (4 \o - 2 ) \phi^{1 - 4 \o} S \cA_\o \cF_\o \,.
\label{genact}
\eea
As stated in footnote 5 from now on we only give the bosonic terms of a supersymmetric Lagrangian unless otherwise stated. The fermionic terms can be read off from the composite formulae (\ref{kinet}).

With these general multiplication and kinetic multiplet construction rules in hand, it is
relative straightforward to construct matter coupled $\cN=1$  Poincar\'e supergravity  models.
Setting $\o = \ft12$ in the action formulae (\ref{genact}) and identifying the general scalar multiplet with the compensating scalar multiplet, we obtain the following Lagrangian describing the coupling of a compensating scalar multiplet to conformal supergravity
\bea
e^{-1} \cL_{\text{SC, R}} &=&  16 \, \phi \Box^c \phi  + S^2  \,.
\label{o0ScalarAction}
\eea
To obtain a Poincar\'e supergravity theory we gauge fix the superconformal transformations
by imposing the following  gauge fixing conditions
\bea
\phi = -\ft12 , \quad \l = 0, \quad b_\m = 0\,,
\label{GaugeFixing}
\eea
where the first choice fixes dilatation, the second fixes the S-supersymmetry and the last one fixes the special conformal transformations. These gauge fixing conditions result into  the following decomposition rules
\bea
\L_{K\m} =  \ft14 \bar\e \phi_\m - \ft14 \bar\eta \p_\m \,,\hskip 1truecm
\eta = \ft12 S \e \,.
\label{DecompostionRules}
\eea
We thus end up with a Poincar\'e multiplet consisting  of a Dreibein $e_\mu{}^a$, a gravitino $\psi_\mu$
and an auxiliary scalar $S$:
 \be
\text{Poincar\'e multiplet:}\hskip .5truecm (e_\m{}^a, \p_\m, S)\,.
 \ee
Substituting the gauge fixing conditions \eqref{GaugeFixing} into the Lagrangian \eqref{o0ScalarAction}, and rescaling with a factor of -2, we obtain the following Lagrangian for this Poincar\'e multiplet (including the fermionic terms):
\bea
e^{-1} \cL_{\text{R}} &=& R - 2 S^2 - \bar\p_\m \g^{\m\n\r} \, \nabla_\n (\o) \p_\r\,.
\label{Poincare}
\eea
This Lagrangian is invariant under the following transformation rules;
\bea
\d e_\m{}^a &=& \ft12 \bar\e \g^a \p_\m \,,\\
\d \p_\m &=& \nabla_\m (\o)\e + \ft12 S \g_\m \e \,,\nn\\
\d S &=& \ft14 \bar\e \g^{\m\n} \p_{\m\n} (\o) - \ft14 S \bar\e \g^\m \p_\m  \,,
\label{OffShellTransformationRules}
\eea
where
\bea
\nabla_\m (\o) \e &=& \Big( \partial_\m + \ft14 \o_\m{}^{ab} \g_{ab} \Big) \e, \quad \p_{\m\n} = \nabla_{[\m}(\o) \p_{\n]}\,.
\eea
This matches with the expressions given in  \cite{Howe:1977us}.
We note that  if we define a torsionful spin-connection as follows
\bea
\O_\m{}^{\pm ab} = \o_\m{}^{ab} + S \e_\m{}^{ab}\,,
\label{torsion}
\eea
the supersymmetry transformations \eqref{OffShellTransformationRules} can be rewritten as \cite{Andringa:2009yc}
\bea
\d e_\m{}^a &=& \ft12 \bar\e \g^a \p_\m \,,\\
\d \p_\m &=& \nabla_\m (\O^-)\e \,,\nn\\
\d S &=& \ft14 \bar\e \g^{\m\n} \p_{\m\n} (\O^-)   \,.
\eea

Finally, a supersymmetric cosmological constant can be added to the Poincar\'e Lagrangian
 \eqref{Poincare} by multiplying  four copies of the compensating multiplet:
\begin{equation}
Z = \phi^{4} \,,\hskip 1truecm
\O = 4 \phi^3 \l  \,,\hskip 1truecm
F = 4 \phi^3 S + 6 \phi^2 \bar\l \l \,.
\label{Weight2CompositeSn}
\end{equation}
Using these expressions into the action formula (\ref{Weight2Action}), and gauge fixing via (\ref{GaugeFixing}) and scaling the resulting Lagrangian with factor of $-2$, we obtain the following supersymmetric cosmological constant Lagrangian (including the fermionic terms)
\bea
e^{-1} \cL_{\text{C}} &=& S + \ft18 \bar\p_\m \g^{\m\n} \p_\n \,,
\label{CosmologicalConstant}
\eea
which can be added to the Poincar\'e Lagrangian \eqref{Poincare}. This finishes our review of the construction of Poincar\'e supergravity with a cosmological constant via superconformal methods.

\section{Supersymmetric Four-Derivative Invariants}\label{s3}

In this section we review the construction of supersymmetric $R^2$ and $R_{\m\n}^2$ invariants by using superconformal techniques and applying various maps between the scalar, Yang-Mills and Poincar\'e multiplets. We consider the following most general expression of mass dimension 4:
\bea
e^{-1} \cL^{(4)} &=& a_1 R_{\m\n} R^{\m\n} + a_2 R^2 + a_3 R S^2 + a_4 S \Box S + a_5 S^4 \,.
\eea
As explained in the introduction, naively one would expect that a proper combination of a Ricci scalar squared and Ricci tensor squared supersymmetric invariant gives rise to the supersymmetric NMG model which should be ghost free. However, a spectrum analysis of $R^2$ and $R_{\m\n}^2$ extended Poincar\'e supergravity around $AdS_3$ reveals that the mere use of off-shell $R^2$ and $R_{\m\n}^2$ invariants is not sufficient to obtain a ghost-free model \cite{Bergshoeff:2010mf}. In particular, to obtain a ghost-free model, one needs a third supersymmetric invariant to  cancel the off-diagonal $RS^2$ term. We will show that, besides the $R^2$ and $R_{\m\n}^2$ invariants, indeed such a third supersymmetric off-diagonal invariant exists. Below we will consider the construction of these three different invariants one after the other. In the last subsection we will apply these results to construct the supersymmetric NMG model \cite{Andringa:2009yc,Bergshoeff:2010mf}.

\subsection{The Supersymmetric $R^2$ Invariant } \label{ss:R2}

In this section, we construct the supersymmetric completion of the $R^2$ term. In order to achieve this, we first need to find  the superconformal map from a neutral scalar multiplet to the Weyl multiplet. There may  be various maps from different scalar multiplets to the Weyl multiplet. However, since the lowest component of the neutral multiplet carries no Weyl weight, it has the special feature that it can be multiplied to another scalar multiplet without changing its properties, but only gives rise to composite expressions, like in (\ref{multip}). This property plays a crucial role   in the construction of higher derivative supergravity invariants later on. Considering a neutral scalar multiplet $(\s, \p, N)$ and two compensating multiplets, $(\phi_1, \l_1, S_1)$ and $(\phi_2, \l_2, S_2)$, sequential use of the inverse multiplication rule (\ref{decomp}) implies the following composite expressions
\bea
\s &=& \phi_1^{-3} S_2 \,,\nn\\
\p &=& - 4 \phi_1^{-3} \slashed{\cD} \l_2 - 3  \phi_1^{-4} S_2 \l_1 \,,\nn\\
N &=& 16 \phi_1^{-3} \Box^C \phi_2 - 3 \phi_1^{-4} S_1 S_2 + 12 \phi_1^{-4} \bar\l_1 \slashed{\cD} \l_2 + 6 \phi_1^{-5} S_2 \bar\l_1 \l_2 \,.
\label{MainMapR2}
\eea
We note that this map reduces to
 \bea
(\s, \quad \p, \quad N ) \quad \longleftrightarrow \quad \Big(S, \quad \g^{\m\n} \p_{\m\n}(\O^-), \quad \widehat{R}(\O^\pm) \Big) \,.
\label{R2Map}
\eea
upon rescaling the map (\ref{MainMapR2}) with an overall $- 1/8$ factor, setting the two compensating multiplets equal to each other and gauge fixing via (\ref{GaugeFixing}). The gauge fixed result matches with the result of \cite{Andringa:2009yc}. Here the torsionful Ricci scalar reads
\bea
\widehat{R}(\O^\pm) &=& R(\o) + 6 S^2 + 2 \bar\p_\m \g_\n \p^{\m\n}(\O^-) + \ft12 S \p_\m \g^{\m\n} \p_\n \,.
\label{TorsionfulRicciScalar}
\eea
 The conformal action for the neutral scalar multiplet  can easily be obtained using the action formula (\ref{genact}). Setting $\o = 0$ in (\ref{genact}), gauge fixing according to (\ref{GaugeFixing}), and multiplying the resulting action with an overall $-1/8$ factor, we obtain the following  action for a Poincar\'e neutral scalar multiplet:
\bea
e^{-1} \cL_{\text{Neutral}} &=& - \partial_\m \s \partial^\m \s  + \ft1{16}  N^2 \,.
\label{w0ScalarAction}
\eea
Using the map (\ref{R2Map}), we obtain the supersymmetric completion of the Ricci scalar squared action \cite{Andringa:2009yc}
\bea
e^{-1} \cL_{R^2} &=& \ft1{16} R^2 - \partial_\m S \partial^\m S + \ft34 R S^2 + \ft94 S^4 \,.
\label{R2Action}
\eea

\subsection{The Supersymmetric $R_{\m\n}^2$ Invariant}\label{ss:Rmn2}

To construct a supersymmetric $R_{\m\n}^2$ invariant we consider a superconformal Yang-Mills multiplet, see table \ref{table1}. After gauge-fixing the superconformal symmetries via eq.~\eqref{GaugeFixing},
one finds the following  map between the Yang-Mills and  supergravity Poincar\'e  multiplets
\cite{Andringa:2009yc}:
\bea
\Big(\O_\m{}^{+ab} , \quad \p^{ab}(\O^-) \Big) \quad \longleftrightarrow \quad (A_\m^I, \quad \vf^I ) \,.
\label{Rmn2Map}
\eea
We see that  the map (\ref{Rmn2Map}) associates the Yang-Mills field strength $F_{ab}^I$  to the torsionful Riemann tensor $R_{ab}{}^{cd}(\O^+)$ given by
\bea
R_{\m\n}{}^{ab}(\O^\pm) &=& R_{\m\n}{}^{ab}(\o) \pm \ft12 S \bar\p_\m \g^{ab} \p_\n \pm 2 \partial_{[\m} S \e_{\n]}{}^{ab} + 2 S^2 e_{[\m}{}^a e_{\n]}{}^b \,.
\label{TorsionfulRiemannTensor}
\eea
Therefore, all we need to do in order to obtain a supersymmetric $R_{\m\n}^2$ invariant is to first construct a superconformal
Yang-Mills action. After substituting the gauge-fixing conditions  (\ref{GaugeFixing}) into this action, one can simply obtain the supersymmetric completion of a $R_{\m\n}^2$ action, modulo a supersymmetric $R^2$ action, by applying the map (\ref{Rmn2Map}), since in three dimensions the Riemann tensor reads
\bea
R_{\m\n ab} &=& \e_{\m\n\r} \e_{abc} G^{\r c} \,,
\label{RiemannExpansion}
\eea
where $G_{\m a}$ is the Einstein tensor.

For the construction of a superconformal Yang-Mills action, we  turn to the action  (\ref{Weight2Action}) for the auxiliary scalar multiplet. Since we are after the supergravity coupled Yang-Mills action, we wish to express the components of the auxiliary multiplet in terms of the
components of the Yang-Mills multiplet thereby using the compensating multiplet to balance the conformal weights.
We thus find the following composite expressions:
\bea
Z &=& \phi^{-2} \bar\vf^I \vf^I \,,\nn\\[.1truecm]
\O &=& - 2 \phi^{-3} \l \bar\vf^I \vf^I - \phi^{-2} \g \cdot \widehat{F}^I \vf^I \,,\nn\\[.1truecm]
F &=& - \phi^{-2} \widehat{F}^I_{\m\n} \widehat{F}^{\m\n I} - 2 \phi^{-3} S \bar\vf^I \vf^I  - 8 \phi^{-2} \bar\vf^I \slashed{\cD} \vf^I \nn\\[.1truecm]
&&-  2 \phi^{-4} \bar\l \g \cdot \widehat{F} \vf^I
 - 3 \phi^{-4} \bar\l \l \bar\vf^I \vf^I\,.
\label{YangMillsScalarComposite}
\eea
Upon substituting these composite expressions into the action  (\ref{Weight2Action}), using the gauge fixing conditions (\ref{GaugeFixing}), and multiplying the resulting action with an overall $-1/16$ factor, we find the following supersymmetric Yang-Mills action (including the fermionic terms)
\bea
e^{-1} \cL_{YM} &=& - \ft14 F_{\m\n}^I F^{\m\n I} - 2 \bar\vf^I \slashed{\nabla} \vf^I + \ft12 \bar\p_\r \g^{\m\n} \g^\r \vf^I F_{\m\n}^I + S \bar\vf^I \vf^I \nn\\
&& - \ft12 \bar\vf^I \vf^I \bar\p_\m \p^\m + \ft18 \bar\vf^I \vf^I \bar\p_\m \g^{\m\n} \p_\n \,.
\label{F2Action}
\eea
Applying  the map (\ref{Rmn2Map}) between the Yang-Mills and Poincar\'e multiplets we obtain the following supersymmetric completion of a $R_{\m\n}^2 + R^2$ action \cite{Andringa:2009yc}:
\bea
e^{-1} \cL_{R_{\m\n}^2 + R^2} &=& - R_{\m\n} R^{\m\n} + \ft14 R^2 + 2 \partial_\m S \partial^\m S - R S^2 - 3 S^4 \,.
\label{Rmn2R2Action}
\eea
Finally, subtracting the $R^2$ part, see eq.~\eqref{R2Action}, we obtain the following $R_{\m\n}^2$ action
\bea
e^{-1} \cL_{R_{\m\n}^2} &=& R_{\m\n} R^{\m\n} - 6 \partial_\m S \partial^\m S +  4 R S^2 + 12 S^4 \,.
\label{RmnAction}
\eea
Before proceeding, we wish  to discuss an alternative  way of constructing the Ricci tensor squared invariant in order to prepare the reader for the construction method of the $R_{\m\n}^3$ invariant in the next section. Since the Weyl multiplet consists of  gauge fields only, the components of the  Weyl multiplet do not show up explicitly in a superconformal Lagrangian. Therefore, alternatively we can start from a rigid higher derivative action for a scalar multiplet with appropriate weight, and then consider the conformal supergravity coupled model.

For the 4-derivative case, we  need a rigid theory that contains a term proportional to $\cA \Box \Box \cA$ since such a term would include the Ricci tensor squared when we couple to conformal gravity and replace this term by $\cA \Box^c \Box^c \cA$. The rigid conformal Lagrangian that includes such a term is given by
\bea
\cL_{\Phi} &=&   \Phi \Box \Box \Phi + \ft1{16} P \Box P \,,
\eea
where $(\Phi, \Psi, P)$ are the fields of a $\o_\Phi = - \ft12$ scalar multiplet. Considering the supergravity coupled model, we obtain the following bosonic terms
\bea
e^{-1} \cL_{\Phi} &=& \Phi \Box^c \Box^c \Phi + \ft1{16}  P \Box^c P  \nn\\[.1truecm]
&=& \Big( R_{\m\n} R^{\m\n} - \ft{23}{64} R^2 \Big)  \Phi^2  + \ft1{16} P \Box P - \ft1{128} R P^2 \nn\\[.1truecm]
&& + \ft32 R \partial_\m \Phi \partial^\m \Phi + \ft1{4} R \Phi \Box \Phi - 4 R_{\m\n} \partial^\m \Phi \partial^\n \Phi + (\Box \Phi)^2 \,.
\label{SCRmnRmn}
\eea
As our supergravity construction is based on a compensating multiplet, we would like to express the $\o_\Phi = -\ft12$ multiplet in terms of a compensating multiplet. Using the multiplication rule (\ref{multip}), we obtain the following expressions:
\bea
\Phi = \phi^{-1}, \quad P = - \phi^{-2} S\,.
\eea
Making this replacement in the Lagrangian \eqref{SCRmnRmn} and imposing the gauge-fixing conditions (\ref{GaugeFixing}), we obtain the supersymmetric completion of the Ricci tensor squared given in \eqref{RmnAction}.

\subsection{Supersymetric $R S^{n}$ Invariants}\label{ss:sn+1}

In this subsection, we construct a third supersymmetric invariant which is a supersymmetric extension of the off-diagonal $RS^2$ term. In fact, we will construct a whole class of
supersymmetric $R S^{n}$ (for any positive integer $n$) invariants. This family of invariants  was already obtained in \cite{Bergshoeff:2010mf} using superspace. Here we give an alternative derivation using components.
For $n=-1$ the invariant coincides with the supersymmetric cosmological constant \eqref{CosmologicalConstant},
while for $n=0$ the invariant is given by the supersymmetric Einstein-Hilbert term, see eq.~\eqref{Poincare}.
For all $n\ge 1$  the invariant does not contain a purely gravitational term.

It is instructive to first construct the $n=2$ invariant  and then derive  the general formula.
Multiplying a neutral multiplet $(\s\,, \p\,, N)$ with an auxiliary multiplet $(Z\,, \Omega\,, F)$ we obtain another
auxiliary multiplet $(Z^\prime\,, \Omega^\prime\,, F^\prime)$ given by
\bea
Z' &=& \s Z \,, \quad  F' = \s F + Z N\,.
\label{2to2Map}
\eea
Substituting these composite expressions  into the action formula \eqref{Weight2Action} we obtain the following action
describing the coupling of a neutral scalar multiplet to an auxiliary scalar multiplet:
\bea
e^{-1} \cL_{\text{AN}} &=& \s F + Z N\,.
\label{L02}
\eea
 We now substitute the composite expressions given in (\ref{Weight2CompositeSn}) and  impose the gauge-fixing conditions (\ref{GaugeFixing}) after which we obtain the following action:
\bea
e^{-1} \cL_{\text{CN}} &=& - \ft12 \s S + \ft1{16} N \,.
\label{L01}
\eea
Finally, applying the map (\ref{R2Map}) in this Lagrangian we obtain
\bea
e^{-1} \cL_{S^2} &=& R - 2 S^2\,,
\eea
which matches with (the bosonic part of) the Poincar\'e supergravity Lagrangian (\ref{Poincare}).

It is now straightforward to  generalize this construction from $n=2$ to general values of $n$.  By sequentially multiplying an auxiliary multiplet with a neutral multiplet, we obtain the general formula
\bea
Z^{(n)} &= \s^n Z\,,  \quad \quad F^{(n)} = \s^n F + n \s^{n-1} Z N \,,
\eea
where  we have only given the terms relevant to the bosonic part of the $RS^n$ action. Using these expressions into the density formula (\ref{Weight2Action}) we obtain the following bosonic terms
\bea
e^{-1} \cL^{(n)}_{\text{AN}} &=& \s^n F + n \s^{n-1} Z N \,.
\label{PreSn}
\eea
 Next, upon substituting  the composite expressions (\ref{Weight2CompositeSn}), imposing
  the gauge fixing conditions (\ref{GaugeFixing}) and applying the map (\ref{R2Map}), we obtain the following Lagrangian:
\bea
e^{-1} \cL_{R S^n} &=& S^{n+2} + \frac{n + 1}{6n - 2} R S^{n} \,.
\label{SnAction}
\eea
This Lagrangian  agrees with the superspace result given in \cite{Bergshoeff:2010mf}.

In the next section we will need the $n=2$ invariant which is given by
\bea
e^{-1} \cL_{RS^2} &=& S^4 + \ft3{10} R S^2 \,.
\label{Need1}
\eea
For future reference we also give the explicit expression of the $n=4$ invariant:
\bea
e^{-1} \cL_{RS^4} &=& S^6 + \ft5{22} R S^4 \,.
\label{Need2}
\eea

\subsection{Supersymmetric  New Massive Gravity}

Given the results obtained in the previous three subsections we can now write the most general supersymmetric terms of mass dimension 4 as follows:
\bea
e^{-1} \cL_{\text{SQuad}} &=& e^{-1} \Big( a_1\cL_{R_{\m\n}^2} + a_2 \cL_{R^2}  + a_3 \cL_{R S^2} \Big) \,,\nn\\[.2truecm]
&=& a_1 R_{\m\n} R^{\m\n} + a_{2} R^2 - ( 16 a_2 + 6a_1 ) \partial_\m S \partial^\m S \nn\\[.2truecm]
&& + \Big( 4a_1 + 12 a_2 + \frac{3}{10} a_3 \Big) R S^2 + ( 12a_1 + 36 a_2 + a_3 ) S^4 \,.
\eea
As discussed in the introduction, we would like to cancel the kinetic term for $S$ and the off-diagonal $R S^2$ term.
This leads to the following two restrictions on the three coefficients $(a_1,a_2,a_3)$:
\be
a_2 = - \ft{3}{8}a_1\,,\hskip 2truecm  a_3 = \ft53 a_1\,.
\ee
Therefore, the terms of mass dimension 4 in  supersymmetric NMG, or shortly SNMG,  are given by
\bea
e^{-1} \cL^{(4)}_{\text{SNMG}} &=& R_{\m\n} R^{\m\n} - \ft38 R^2 + \ft16 S^4 \,.
\label{SNMG}
\eea

This concludes our discussion of how to construct a ghost-free supersymmetric NMG model.

\section{Supersymmetric Six-Derivative Invariants}\label{s4}

In this section we push the analysis of the previous section one level further and consider the
$\cN=1$ supersymmetrization of the following terms with mass dimension 6:\,\footnote{These are the most general terms of mass dimension 6 except for the terms
$ R_{\m\n} \Box R^{\m\n}\,,  R \Box R \,, S \e_{\m\n\r} R^{\m}{}_{\s} \nabla^\r R^{\n\s}$
and $(\Box S)^2$.  We do not consider the first two terms since they lead to models with two massive gravitons one of
which is a ghost. It turns out that the last two terms are only needed for  the supersymmetrization of the first two terms.}
\bea\label{mostg6}
e^{-1} \cL^{(6)} &=& a_1 R_{\m\n} R^{\n\r} R_\r{}^\m  + a_2 R^3 + a_3 R R_{\m\n} R^{\m\n} +  a_4 R^2 S^2 + \nn\\[.1truecm]
&&   + a_5\, R S^4 + a_6 S^2 R_{\m\n} R^{\m\n} + a_7 R_{\m\n} \partial^\m S \partial^\n S +  \nn\\[.1truecm]
&&  + a_{8}\, R \partial_\m S \partial^\m S + a_{9} R S \Box S + a_{10} S^2 \partial_\m S \partial^\m S + a_{11} S^6 \,.      \eea
Here $a_1\,\dots,a_{11}$ are arbitrary coefficients. Requiring supersymmetry will lead to relations between these coefficients leaving us with as many parameters as supersymmetric invariants we construct in this section.  As discussed in the introduction, requiring a ghost-free model at the linearized level leads to several restrictions on the remaining independent parameters.
These restrictions  will be discussed in the next section.

In this section we will merely focus on the construction of the different supersymmetric invariants.
In the first subsection we will construct the supersymmetrization of the three  purely gravitational terms
in the Lagrangian \eqref{mostg6}, corresponding to the coefficients $a_1,\dots,a_3$. For completeness,
we will also construct the supersymmetric $R\Box R$ and $R_{\m\n}\Box R^{\m\n}$ invariants.
In the next subsection we will construct the supersymmetric completion of five of the off-diagonal terms in \eqref{mostg6} corresponding to the
coefficients $a_4,a_6,a_7,a_{9}$.
The supersymmetric completion of the $a_5$-term was already given in eq.~\eqref{Need2}.


\subsection{Diagonal Invariants}

In this subsection we will discuss, one after the other,  how to obtain the supersymmetric extensions g1,...,g5 of the first three purely gravitational terms in (4.1) [g1, g2, g3], together with the supersymmetric extensions of the $R\Box R$ and $R_{\m\n}\Box R^{\m\n}$ terms [g4, g5].

\subsection*{g1: Supersymmetric $R^3$ Invariant}

 The construction of the supersymmetric $R^3$ invariant  requires the introduction of a second neutral multiplet
 $(\s', \p' ,N')$.
 Using the multiplication rule (\ref{multip}), we first construct a new auxiliary multiplet $(Z^\prime,\Omega',F^\prime)$ in terms of an auxiliary multiplet $(Z,\Omega,F)$ and this second  neutral multiplet $(\s', \p', N')$:
\bea
Z' &=& \s'^2 Z, \qquad  F' = \s'^2 F + 2 \s' Z N' \,.
\label{2to2Map}
\eea
 We now use  the fact the fields of the auxiliary multiplet $(Z,\Omega,F)$ can be expressed in terms of the components of  the compensating multiplet as
 \be\label{subs}
 Z = \phi S\,,\hskip 2truecm F = 16 \phi \Box^c \phi + S^2
  \ee
by  first multiplying a compensating multiplet with a curvature  multiplet and, next, substituting the composite expressions (\ref{CurvtoComp}) for the components of the curvature multiplet. Next, we use the action formula for $(Z^\prime,\Omega',F^\prime)$ where we write the components as in eq.~\eqref{2to2Map} with the substitutions
\eqref{subs} being made. This
gives rise to the following action that describes the coupling of a neutral multiplet to a compensating multiplet
\bea
e^{-1} \cL_{\text{CN}} &=& 16 \, \s'^2 \phi \Box \phi - 2 R \phi^2 \s'^2 + S^2 \s'^2 + 2\s' \phi S N' \,.
\label{CompToNat}
\eea

 The final step of the construction requires one to express $\sigma'$ in terms of the Ricci scalar.
 To achieve this we make use of the map (\ref{MainMapR2}) that expresses the components of the neutral scalar multiplet
 as the product of two compensating  multiplets $(\phi_1, \l_1, S_1)$ and $(\phi_2, \l_2, S_2)$.
  Using the multiplication rule, we can write the components $(\phi_2, \l_2, S_2)$ as
\bea
\phi_2 = \s \phi_1 \,,\qquad \l_2 = \phi_1 \p + \s \l_1\,, \qquad S_2 = \s S_1 + \phi_1 N + \bar\l_1 \p \,.
\eea
Then, using these expressions in (\ref{MainMapR2}) and setting $\phi_1 = \phi$, we obtain the following bosonic expressions
\bea
\s' &=& \phi^{-2} N + \phi^{-3} \s S \,,\nn\\
N' &=& 16 \phi^{-3} \Box^C (\phi \s) - 3 \phi^{-4} \s S^2 - 3 \phi^{-3} S N \,.
\label{MComp}
\eea
Next, applying the map (\ref{MainMapR2}) in the above expressions  and gauge fixing according to (\ref{GaugeFixing}), we indeed have expressed
$\sigma'$ in terms of the Ricci scalar. Finally, substituting the resulting expressions into the action (\ref{CompToNat})
we find that the bosonic part of  the supersymmetric completion of the $R^3$ term
is given by
\bea
e^{-1} \cL_{R^3} &=& R^3 + 14 R^2 S^2 + 80 R S^4 + 32 R S \Box S - 384 S^2 \partial_\m S \partial^\m S + 160 S^6 \,.
\label{R3}
\eea

\subsection*{g2: Supersymmetric $R R_{\m\n} R^{\m\n}$ Invariant}

We next consider the supersymmetric completion of the $R R_{\m\n} R^{\m\n}$ term. In order to do this
we first wish to construct a supersymmetric $\sigma F_{\mu\nu}^I F^{\mu\nu I}$ term and next apply the maps (\ref{R2Map}) and (\ref{Rmn2Map}).
To achieve this we first multiply the auxiliary multiplet with a neutral multiplet such that we obtain a new auxiliary multiplet. The bosonic composite expressions for  the new auxiliary multiplet $\big(Z^\prime, \Omega^\prime, F^\prime\big)$ are given by
\begin{equation}
Z' = \s Z, \quad F'= \sigma F + Z N\,.\label{Fonly}
\end{equation}
Plugging this into the action formula \eqref{Weight2Action} we obtain the following  action:
\begin{equation}\label{boscons}
e^{-1}{\cal L}_{\text{AN}} = \sigma F + Z N\,.
\end{equation}
Next, we substitute the composite expressions \eqref{YangMillsScalarComposite} to obtain the action
\bea
e^{-1} \cL_{\s F_{\m\n}^2} &=& - \ft14 \s F_{\m\n}^I  F^{\m\n I}\,,
\label{SnFF}
\eea
where we have gauge fixed according to (\ref{GaugeFixing}).
Finally, using the composite expressions (\ref{MComp}), and applying  the maps (\ref{R2Map}) and (\ref{Rmn2Map}), we obtain the following supersymmetric invariant:
\bea
e^{-1} \cL_{R^3 + R R_{\m\n} R^{\m\n}}
&=& R R_{\m\n} R^{\m\n} - \ft14 R^3 - 2 R \partial_\m S \partial^\m S + 4 S^2 R_{\m\n} R^{\m\n} + 7 R S^4  \nn\\
&& - 8 S^2 \partial_\m S \partial^\m S + 12 S^6 \,.
\eea
Combining this action with the supersymmetric $R^3$ action, see eq.~\eqref{R3}, we find that the  supersymmetric completion of the $R R_{\m\n} R^{\m\n}$ invariant is given by:
\bea
e^{-1} \cL_{R R_{\m\n} R^{\m\n}} &=& R  R_{\m\n} R^{\m\n} + 4 S^2 R_{\m\n} R^{\m\n} - 2 R \partial_\m S \partial^\m S + 8 R S \Box S \nn\\
&& + \ft72 R^2 S^2 + 27 R S^4 - 104 S^2 \partial_\m S \partial^\m S + 52 S^6 \,.
\eea
\subsection*{g3: Supersymmetric $R_{\m\n}^3$ Invariant}

We now discuss the construction of the supersymmetric Ricci tensor cube invariant. Unlike the other invariants, the $R_{\m\n}^3$ cannot cannot be obtained by applying a map between a scalar or  Yang-Mills multiplet and the Weyl multiplet. Instead, we  look for a K-invariant conformal quantity that gives rise to a $R_{\m\n}^3$ action. Given the following K-transformations
\bea
\d_K \Box^c \cA &=& - 2 (2 \o_\cA  - 1 ) \L_K^a \cD_a \cA \,,\nn\\
\d_K \Box^c \Box^c \cA &=&-4 (2\o_\cA + 1) \L_K^a \cD_a \Box^c \cA \,,
\eea
we observe that the $\phi \Box^c \phi$ and $\Phi \Box^c \Box^c \Phi$ terms, which gave rise to a supersymmetric Poincar\'e and $R_{\m\n}^2$ action arose naturally by demanding  K-invariance and by requiring the conformal weight to match the required  factor $+3$.  A natural generalization of this construction is to consider
\bea
\d_K \Box^c \Box^c \Box^c \cA &=& - 6 (2\o_\cA + 3 ) \L_K^a \cD_a \Box^c \Box^c \cA \,.
\eea
This transformation rule shows that given a scalar multiplet $(\tau, \zeta, L)$ with the conformal weight of the lowest component given by $ \o_\tau = -\ft32$, the following Lagrangian gives rise to a supersymmetric completion of the $R_{\m\n}^3$ term:
\bea
e^{-1} \cL_\t &=&  \t \Box^c \Box^c \Box^c \t + \ft1{16} L \Box^c \Box^c L \nn\\[.1truecm]
&=& \Big( \ft32R_{\m\n}\partial^\m \partial^\n R - \ft{39}{64} R \Box R + 24 R_{\m\n} R^{\n\r} R_{\r}{}^\m - \ft{45}{2} R R_{\m\n} R^{\m\n} + \ft{2463}{512} R^3 \Big) \t^2 \nn\\[.1truecm]
&& + \ft1{16} \Big(R_{\m\n} R^{\m\n} - \ft{23}{32} R^2 \Big) L^2 + \ft3{32} R \partial_\m L \partial^\m L + \ft1{64} R L \Box L\nn\\[.1truecm]
&&  - \ft14 R_{\m\n} \partial^\m L \partial^\n L + \ft1{16} (\Box L)^2  \,.
\eea
Since our supergravity construction is based on a compensating multiplet, we would like to express the $\o_\t = -\ft32$ multiplet in terms of the compensating multiplet. Using the multiplication rule (\ref{multip}), we obtain the following expressions:
\bea
\t = \phi^{-3}, \quad L = - 3 \phi^{-4} S \,.
\eea
Making this  replacement in the above action and gauge fixing according to (\ref{GaugeFixing}), we obtain the following supersymmetric completion of the Ricci tensor cube term:
\bea
e^{-1} \cL'_{R_{\m\n}^3} &=& \ft3{64} R \Box R + 8 R_{\m\n} R^{\n\r} R_{\r}{}^\m - \ft{15}2 R R_{\m\n} R^{\m\n} + \ft{821}{512} R^3 \nn\\[.1truecm]
&& + \ft34 R_{\m\n} R^{\m\n} S^2 - \ft{69}{256} R^2 S^2  + \ft{9}{8} R \partial_\m S \partial^\m S  - 3 R^{\m\n} \partial_\m S \partial_\n S   \nn\\[.1truecm]
&&+ \ft3{16} R S \Box S + \ft3{4} (\Box S)^2 \,.
\eea
Subtracting the $R^3, R R_{\m\n} R^{\m\n}$ invariants and the $R \Box R$ invariant, which we shall construct  below, see eq.~\eqref{RBoxR}, the final expression of the $R_{\m\n}^3$ action reads
\bea
e^{-1} \cL_{R_{\m\n}^3} &=& R^{\m\n} {R}_{\m}{}^{\r} {R}_{\n\r} + \ft{123}{32}\, {S}^{2} {R}^{\m\n} {R}_{\m\n} + \ft{223}{512}\, R^{2} S^{2} - \ft{117}{64} R {\partial}^{\m}{S}\,  {\partial}_{\m}{S}\,  \nn\\[.1truecm]
&& + \ft{31}{32} R S \Box S - \ft{3}{8}\, {R}^{\m\n} {\partial}_{\m}{S}\,  {\partial}_{\n}{S}\,  - \ft{309}{16}\, S^{2} {\partial}^{\m}{S}\,  {\partial}_{\m}{S}\,   + \ft{2357}{256}\, R S^{4} \nn\\[.1truecm]
&& + \ft{527}{32}\, S^{6} \,.
\eea

\subsection*{g4: Supersymmetric $R\Box R$ Invariant}

To construct the $R_{\m\n}^3$ invariant above  we needed to know the result for the $R\Box R$ invariant. We will derive this expression here.
To construct the supersymmetric completion of a $R \Box R$ term we make use of the following observation. When we constructed the supersymmetric $R^2$ invariant in section \ref{ss:R2}, we started from a two-derivative
 action for a neutral scalar multiplet. Next, after gauge fixing the superconformal symmetries,
  we obtained the supersymmetric $R^2$ invariant by making use of the  map \eqref{R2Map}
  between the neutral scalar multiplet and the Weyl multiplet in which the auxiliary field of the neutral scalar multiplet, $N$, is mapped to the torsionful Ricci scalar.
 This suggests that, to obtain a supersymmetric $R\Box R$ term, we should start from a four derivative
  action for the same scalar multiplet  such that the auxiliary scalar $N$ occurs as a $N\Box N$ term. Under the map (\ref{R2Map}), this term will give rise to the desired $R\Box R$ term.

 The problem of constructing a $R\Box R$ invariant is now reduced to constructing a four-derivative action for a neutral scalar multiplet.
Such an action can easily be constructed by using the composite expressions (\ref{MComp}) into the action for the neutral scalar multiplet. Gauge fixing via (\ref{GaugeFixing}) and using the gauge fixed expressions for (\ref{MComp}) into the action (\ref{w0ScalarAction}) we obtain the following supersymmetric completion of the $R \Box R$ term:
\bea\label{RBoxR}
e^{-1} \cL_{R \Box R} &=&  R \Box R + 24 RS \Box S + 16 R \partial_\m S \partial^\m S - 208 S^2 \partial_\m S \partial^\m S + 16 (\Box S)^2 \nn\\
&& + R^2 S^2 + 12 R S^4 + 36 S^6 \,.
\eea

\subsection*{g5: Supersymmetric $R_{\m\n} \Box R^{\m\n}$ Invariant}

For completeness,   we construct the supersymmetric completion of a $R_{\m\n} \Box R^{\m\n}$ term. In the same way that we used in subsection \ref{ss:Rmn2} the gauge-fixed supersymmetric $F_{\m\n}  F^{\m\n}$ action and the map (\ref{Rmn2Map}) to
construct a supersymmetric $R_{\m\n} R^{\m\n}$ invariant we expect that  a supersymmetric $R_{\m\n} \Box R^{\m\n}$ action can be constructed from applying (\ref{Rmn2Map}) to a supersymmetric $F_{\m\n} \Box F^{\m\n}$ action. Our main task is to construct such an action.

Our starting point is the gauge-fixed supersymmetric Yang-Mills action (\ref{F2Action}) with  gauge vector $C_\mu^I$
and curvature $G_{\mu\nu}^I$. We observe that  upon writing
\begin{equation}\label{use}
G_{\m\n}^I = \e_{\m\n\r} \nabla_\s F^{\r\s I}\,,
\end{equation}
the leading $G_{\m\n}^I G^{\m\n I}$ term gets mapped to a higher-derivative
$F_{\m\n}^I \Box F^{\m\n I}$ term. Therefore, using \eqref{use}, we wish to map a conformal Yang-Mills multiplet $(A^I_\m, \vf^I)$ to
another conformal Yang-Mills multiplet $(C^I_\m, \O^I)$.  We find that this is achieved by the following map:
\bea\label{mapV}
\O^I &=& \phi^{-2} \slashed{\cD} \vf^I + \ft18 \phi^{-3} \g \cdot \widehat{F}^I \l - \phi^{-3} \slashed{\cD}\phi \vf^I + \ft14 \phi^{-3} S \vf^I - \ft38 \phi^{-4} \vf^I \bar\l \l \,,\nn\\
\widehat{G}_{\m\n}^I &=&  \e_{\m\n\r} \cD_\s (\phi^{-2} \widehat{F}^{\r\s I} )  \,.
\eea
Applying this map we find that, after gauge-fixing,  the supersymmetric completion of a $F_{\m\n} \Box F^{\m\n}$ term does
not require any other bosonic term:
\bea
e^{-1} \cL_{F_{\m\n} \Box F^{\m\n}} &=& \nabla_\m F^{\m \n I} \nabla^\r F_{\n\r}^I \,.
\label{FBoxF}
\eea
Next, applying the map \eqref{Rmn2Map} between a Yang-Mills multiplet and the Weyl multiplet, we obtain the following action
\bea
e^{-1} \cL'_{R_{\m\n} \Box R^{\m\n}} &=& R_{\m\n} \Box R^{\m\n} - \ft14 R \Box R - 3 R_{\m\n} R^{\n\r} R_{\r}{}^\m  + \ft52 R R_{\m\n} R^{\m\n}  \nn\\ [.1truecm]
&&- \ft12 R^3- 14 R S \Box S - 14 R \partial_\m S \partial^\m S  - 16 \e^{\m\r\s} S R_\m{}^\n \nabla_{\r} R_{\n\s} \nn\\ [.1truecm]
&& + 16 R_{\m\n} \partial^\m S \partial^\n S  + 2 (\Box S)^2 + 24 S^2 \partial_\m S \partial^\m S  \nn\\ [.1truecm]
&&  -16 R^2 S^2 + 64 S^2 R_{\m\n} R^{\m\n} + 64 R S^4 + 192 S^6  \,.
\eea
Finally, subtracting the $R_{\m\n}^3, R R_{\m\n} R^{\m\n}, R^3$ and $R \Box R$ invariants, we obtain the desired supersymmetric completion of the $R_{\m\n} \Box R^{\m\n}$ invariant
\bea
e^{-1} \cL_{R_{\m\n} \Box R^{\m\n}} &=& R_{\m\n} \Box R^{\m\n} + \ft{2097}{32} S^2 R_{\m\n} R^{\m\n} - \ft{8291}{512} R^2 S^2 + \ft{17183}{256} R S^4   \nn\\ [.1truecm]
&& - \ft{671}{64} R \partial_\m S \partial^\m S - \ft{291}{32} R S \Box S + \ft{119}{8} R_{\m\n} \partial^\m S \partial^\n S  \\ [.1truecm]
&& - 16 \e^{\m\r\s} S R_\m{}^\n \nabla_{\r} R_{\n\s}  - \ft{287}{16} S^2 \partial_\m S \partial^\m S + 6 (\Box S)^2 + \ft{6413}{32} S^6 \,.\nonumber
\eea

\subsection{Off-Diagonal Invariants}

In this subsection we will construct, one after the other, the supersymmetric invariants o1\, ...,o4, corresponding to the off-diagonal $a_5,a_6,a_7,a_9$ terms in (4.1). For completeness, we will also give the supersymmetric extension of the off-diagonal term $\epsilon^{\m\n\r}SR_\m{}^\s\nabla_\n R_{\s\r}$ which will not be needed since it only occurs in the $R_{\m\n}\Box R^{\m\n}$ invariant. The 5th off-diagonal invariant  that will be needed is the supersymmetric completion of the $RS^4$ term which was already given in eq.~\eqref{Need2}.

\subsection*{o1: Supersymmetric $S^{2n}R^2$ Invariants}

In the previous subsection we have shown that the supersymmetric cosmological constant and  Poincar\'e supergravity are special cases of a more general supersymmetric $RS^{n}$ invariant  for $n = -1$ and $ n =0$. Here
we will show that similarly the $R^2$ Lagrangian \eqref{R2Action} can be seen as a special case of a more general
  supersymmetric $S^{2n}R^2$ invariant for $n=0$.
  In order to do so, it is instructive to first re-consider the $n=0$ case. We  start from the superconformal scalar multiplet Lagrangian (\ref{o0ScalarAction})
  for a compensating scalar multiplet. Next, using the multiplication rule (\ref{multip}), we replace $\phi$ with $\sigma \phi$ and $S$ with $(\sigma S + \phi N)$. This leads to the following Lagrangian:
\bea
e^{-1} \cL_{\text{SC}} &=& 16 (\s \phi) \Box (\s \phi) - 2 R \phi^2 \s^2 + (\s S + \phi N)^2 \,.
\eea
Substituting the superconformal gauge fixing conditions  (\ref{GaugeFixing}) and applying (\ref{R2Map}) we obtain
\bea
e^{-1} \cL'_{R^2} &=& 4 S \Box S + 4 S^2 + \ft32 R S^2 + \ft14 R^2 \,.
\label{RS2}
\eea
Combining this Lagrangian  with the $S^4$ action given by (\ref{SnAction}) for $n=3$ we recover the $R^2$
Lagrangian \eqref{R2Action} in the form we derived earlier:
\bea
e^{-1}\cL_{R^2} = e^{-1} (\cL'_{R^2} + 5 \cL_{S^4}) = 4 S \Box S + 9 S^4 + 3 R S^2 + \ft14 R^2 \,.
\eea

We next consider the general case and take a compensating scalar multiplet coupled to $n$ copies of a neutral scalar multiplet. In that case we replace in the superconformal scalar multiplet Lagrangian
(\ref{o0ScalarAction})
$\phi$ with $\s^n \phi$ and $S$ with $\s^n S + n \s^{n-1} \phi N$ and obtain:
\bea
e^{-1} \cL_{\text{C} \text{N}^n} &=& 16 (\s^n \phi) \Box (\s^n \phi) - 2 R \s^{2n} \phi^2 + ( \s^n S + n \s^{n-1} \phi N)^2 \,.
\eea
Finally, substituting the superconformal gauge-fixing conditions  (\ref{GaugeFixing}) and applying \eqref{R2Map}  we obtain
\bea
e^{-1} \cL_{ S^{2n}R^2} &=& - 4 (n+1)^2 S^{2n} \partial_\m S \partial^\m S  + \ft14 (n+1)^2 R^2 S^{2n} + (3n + 2)^2 S^{2n + 4}  \nn\\
&& + (3n^2 + 5n + \ft32) R S^{2n + 2} \,,
\eea
 which reduces for $n=0$  to the supersymmetric invariant given in (\ref{RS2}). For simplicity, in the off-diagonal case,
 we will use a basis where terms that belong to other off-diagonal invariants, like the last term in the above Lagrangian,
 have not been canceled. We will only do this for the purely gravitational terms.

\subsection*{o2: Supersymmetric $S^n R_{\mu\nu} R^{\m\n}$ Invariants}\label{ss: SRR}

The Ricci tensor  squared Lagrangian \eqref{Rmn2R2Action}  can also be obtained as a special case of a more general supersymmetric $S^n R_{\m\n}^2$ Lagrangian by setting $n=0$. The construction  is based on the superconformal action (\ref{PreSn}). Substituting
 the composite expressions (\ref{YangMillsScalarComposite})  into (\ref{PreSn}) and gauge-fixing according to
  (\ref{GaugeFixing})
  we obtain the following Lagrangian
\bea
e^{-1} \cL_{\s^n F_{\m\n}^2} &=& - \ft14 \s^n F_{\m\n}^I F^{\m\n I}\,.
\label{SnFF}
\eea
Next, after applying (\ref{R2Map}) and (\ref{Rmn2Map}) we obtain the  Lagrangian:
\bea
e^{-1} \cL_{S^{n} R_{\m\n}^2} &=& - S^n R_{\m\n} R^{\m\n} + \ft14 S^n R^2 + 2 S^n  \partial_\m S \partial^\m S - R S^{n+2} - 3 S^{n+4}\,.
\eea

\subsection*{o3: Supersymmetric $R_{\m\n} \partial^\m S \partial^\n S$ Invariant}

The construction of the  $R_{\m\n} \partial^\m S \partial^\n S$ invariant is based on the superconformal action (\ref{SCRmnRmn}). Using the multiplication rule, we replace the  $\o_\Phi = - \ft12$ multiplet with another $\o_\Phi = - \ft12$, multiplet, i.e., $\Phi \rightarrow \s \Phi$ and $P \rightarrow \s P + \Phi N$. After gauge fixing, the resulting action reads
\bea
e^{-1} \cL'_{R_{\m\n} \partial^\m S \partial^\n S} &=& 4 R_{\m\n} R^{\m\n} S^2 - \ft{31}{16} R^2 S^2 - 64 S^2 \partial_\m S \partial^\m S + 9 R S \Box S + 14 R \partial_\m S \partial^\m S  \nn\\
&& + \ft14 R \Box R - 2 R S^4 - \ft1{32} R^3 - 16 R_{\m\n} \partial^\m S \partial^\n S + 4 (\Box S)^2 \,.
\eea
Subtracting the $R^3$ and $R \Box R$ invariants, we find that the $R_{\m\n} \partial^\m S \partial^\n S$ invariant
is given by
\bea
e^{-1} \cL_{R_{\m\n} \partial^\m S \partial^\n S} &=& R_{\m\n} \partial^\m S \partial^\n S  - \ft58 R \partial_\m S \partial^\m S - \ft14 R S \Box S - \ft14 S^2 R_{\m\n} R^{\m\n} \nn\\
&&  + \ft7{64} R^2 S^2 + \ft5{32} R S^4 + \ft32 S^2 \partial_\m S \partial^\m S + \ft14 S^6 \,.
\eea

\subsection*{o4: Supersymmetric $R S \Box S$ Invariant} \label{ss:RSBoxS}

For the construction of the supersymmetric $R S \Box S$ invariant, our starting point is the action for a neutral scalar multiplet $\big(\s', \p', N'\big)$ coupled to an auxiliary scalar multiplet $\big (Z,\Omega,F\big)$, see eq.~(\ref{L02}):
\bea
e^{-1} \cL_{\text{AN}} &=& \s' F + Z N' \,.
\eea
Following the same strategy that was used when we constructed the  Ricci scalar squared invariant, this action can be cast into the following form:
\bea
e^{-1} \cL_{N^2} &=& \s' \Big( \partial_\m \s \partial^\m \s - \ft1{16} N^2 \Big) \,.
\eea
Next, using the maps (\ref{R2Map}) and (\ref{MComp}), we obtain the following $R^3$ action
\bea
e^{-1} \cL'_{R^3} &=& R^3 + 16 R^2 S^2 + 84 R S^4 - 16 R \partial_\m S \partial^\m S - 64 S^2 \partial_\m S \partial^\m S + 144 S^6 \,.
\eea
Finally, removing the $R^3$ invariant found in (\ref{R3}), we find that the  $R S \Box S$ invariant is given by
\bea
e^{-1} \cL_{R S \Box S} &=& R S \Box S + \ft12 R \partial_\m S \partial^\m S - \ft1{16} R^2 S^2 - \ft18 R S^4 - 10 S^2 \partial_\m S \partial^\m S + \ft12 S^6 \,.
\eea

\subsection*{o5: Supersymmetric $S^n \epsilon_{\m\s\l} R^{\m}{}_{\n} \nabla^{\s} R^{\l\n}$ Invariant}

In this section, we consider the supersymmetrization of the $S^n \epsilon_{\m\s\l} R^{\m}{}_{\n} \nabla^{\s} R^{\l\n}$
term.
The $n=0$ case corresponds to the purely gravitational parity-odd term  $\epsilon_{\m\n\r} R^{\m}{}_{\s} \nabla^{\n} R^{\r\s}$. Our starting point for the construction procedure is the Lagrangian that describes the coupling of two Yang-Mills multiplets to $n$ number of neutral multiplets
\bea
e^{-1} \cL_{2YM} &=& - \ft14 \s^n F_{\m\n}^I G^{\m\n I} \,,
\eea
where the curvature $G_{\m\n}^I$ is the curvature of the gauge vector $C_\mu^I$. We observe that once again,  upon writing $G_{\m\n}^I = \e_{\m\n\r} \nabla_\s F^{\r\s I}$, the leading $G_{\m\n}^I F^{\m\n I}$ term gets mapped to a higher-derivative
$\epsilon_{\m\n\r} F^{\m}{}_{\s}{}^I \nabla_{\n} F^{\r\s I}$ term, which leads to the desired supersymmetric higher derivative action via the Yang-Mills to Poincar\'e multiplet map. Therefore, using (\ref{mapV}) we obtain the following supersymmetric completion of a $\s^n \epsilon_{\m\n\r} F^{\m}{}_{\s}{}^I \nabla_{\n} F^{\r\s I}$ term:
\bea
e^{-1} \cL_{ \s^n \epsilon_{\m\n\r} F^{\m}{}_{\s}{}^I \nabla^{\n} F^{\r\s I}} &=& \s^n \epsilon_{\m\n\r} F^{\m}{}_{\s}{}^I \nabla^{\n} F^{\r\s I} \,.
\eea
Finally, using the neutral multiplet to Poincar\'e map (\ref{R2Map}) and the Yang-Mills to Poincar\'e multiplet map (\ref{Rmn2Map}), we obtain
\bea
e^{-1}\cL_{S^n \epsilon_{\m\n\r} R^{\m}{}_{\s} \nabla^{\n} R^{\r\s}} &=& S^n \epsilon_{\m\n\r} R^{\m}{}_{\s} \nabla^{\n} R^{\r\s} - n S^{n - 1} R_{\m\n} \partial^\m S \partial^\n S +  n S^{n-1} R \partial_\m S \partial^\m S \nn\\
&& + R S^n \Box S - 2 ( n+ 12 ) S^{n + 1} \partial_\m S \partial^\m S + 8 S^{n+1} R_{\m\n} R^{\m\n} \nn\\
&& - 2 R^2 S^{n+1} + 8 R S^{n+3} + 24 S^{n+5} \,.
\eea

\section{Towards 3D Born-Infeld Supergravity}\label{ss:5}

In this section, we use the results we obtained in the previous section to construct a supersymmetric ghost-free
higher-derivative gravity model. First, in the next subsection we
construct the supersymmetric cubic extended NMG model. Next, in a separate subsection, we use this result to propose  an expression for the bosonic part of the supersymmetric 3D Born-Infeld gravity model. In particular, we will show that
different truncations of a perturbative expansion of this expression gives rise to the supersymmetric NMG model  and the supersymmetric cubic extended NMG model.

\subsection{Supersymmetric Cubic Extended NMG} \label{ss:SCENMG}

Given the supersymmetric invariants we constructed in the previous section we are now in a position to calculate the $\cN=1$ supersymmetric completion of cubic extended NMG.
To avoid ghostlike massive gravitons, we restrict ourselves to the use of curvature terms without derivatives, hence we do not consider $R \Box R$ and $R_{\m\n} \Box R^{\m\n}$ terms.
In this case, the most general supersymmetric invariant of mass dimension 6  is given by
\bea\label{mostgcubic}
e^{-1} \cL^{(6)} &=& e^{-1} \Big(a_1 \cL_{R_{\m\n}^3} + a_2 \cL_{R R_{\m\n} R^{\m\n}} + a_3 \cL_{R^3} +  a_4 \cL_{R^2 S^2} + a_5 \cL_{S^2 R_{\m\n}^2} \nn\\
&& \qquad +  a_6 \cL_{R_{\m\n} \partial^\m S \partial^\n S}  +  a_7 \cL_{R S \Box S} + a_8 \cL_{R S^4}  \Big) \nn\\[.2truecm]
&=& a_1 R_{\m\n} R^{\n\r} R_\r{}^\m  +a_2 R R_{\m\n} R^{\m\n} + a_3 R^3  \nn\\
&&  + \Big( - \ft{3}{8}a_1 + a_6 \Big) R_{\m\n} \partial^\m S \partial^\n S + \Big( \ft{31}{32}a_1 + 8 a_2 + 32 a_3 - \ft14 a_6 + a_7 \Big) R S \Box S  \nn\\
&& + \Big( \ft{123}{32} a_1 + 4 a_2 - a_5 - \ft14 a_6 \Big) S^2 R_{\m\n} R^{\m\n} + \Big( - \ft{117}{64}a_1 - 2 a_2 - \ft58 a_6 + \ft 12 a_7  \Big) R \partial_\m S \partial^\m S  \nn\\
&& + \Big( \ft{223}{512}a_1 + \ft72 a_2 + 14 a_3 + a_4 + \ft14 a_5 + \ft7{64} a_6 - \ft1{16} a_7 \Big) R^2 S^2 \nn\\
&& + \Big( - \ft{309}{16}a_1 - 104 a_2 - 384 a_3 - 16 a_4 + 2 a_5 + \ft32 a_6 - 10 a_7 \Big) S^2 \partial_\m S \partial^\m S \nn\\
&& + \Big( \ft{2357}{256}a_1 + 27 a_2 + 80 a_3 + \ft{19}{2} a_4 - a_5 + \ft5{32} a_6 - \ft{1}{8} a_7 + \ft{5}{22} a_8 \Big)  R S^4 \nn\\
&&   + \Big( \ft{527}{32}a_1 + 52 a_2 + 160 a_3 + 25 a_4 - 3 a_5 + \ft14 a_6 + \ft12 a_7 + a_8 \Big) S^6 \,.
\label{GenCubAct}
\eea
corresponding to the 8 supersymmetric invariants we have constructed.
The requirement that  all off-diagonal terms and the term $S^2\partial_\m S\partial^\m S$ should vanish to
avoid ghosts when linearizing around an $\text{AdS}_3$ background leads to  7 relations between the 8 coefficients. Setting $a_1=1$ we find the following unique solution:
\bea
\hskip -.2truecm  a_2 = - \ft98, \quad a_3 = \ft{17}{64}, \quad a_4 = - \ft3{32}, \quad a_5 = - \ft34 , \quad a_6 = \ft38 , \quad a_7 = - \ft38,  \quad a_8 = - \ft{33}{160} \,.
\eea
 Substituting these values into the action (\ref{GenCubAct}), we find that the terms of mass dimension 6 in the supersymmetric cubic extended new massive gravity model, shortly called the SCNMG model,  are given by
\bea
e^{-1} \cL^{(6)}_{\text{SCNMG}} &=& R_{\m\n} R^{\n\r} R_\r{}^{\m} - \ft98 R R_{\m\n} R^{\m\n} + \ft{17}{64} R^3 + \ft3{40} S^6 \,.
\label{SuperCNMG}
\eea
The fact that supersymmetry and the absence of ghosts at the linearized level uniquely leads to the expression \eqref{SuperCNMG} is
 one of the main results of this work.

\subsection{Towards 3D Born-Infeld Supergravity}

An example of a specific 3D higher-derivative gravity theory with an infinite number of higher-derivative terms
is the Born-Infeld gravity theory
\bea
I_{\text{BI}} (G_{\m\n}) &=& 4 m^2 \int d^3 x  \sqrt{-g} - 4 m^2 \int d^3 x  \sqrt{ - \det (g_{\m\n} - \ft1{m^2} G_{\m\n} ) }  \,,
\label{MBIM}
\eea
where $G_{\m\n}$ is the Einstein tensor and $m$ is a mass parameter.
It was observed in \cite{Gullu:2010pc} that upon making a perturbative expansion in $1/m$ and keeping only terms linear and quadratic in the Einstein tensor one ends up with the ghost-free NMG model \cite{Bergshoeff:2009hq}. Furthermore, keeping also the terms of mass dimension six  in the Einstein tensor  one ends up with the  six-derivative cubic extended  NMG model \cite{Gullu:2010pc,Sinha:2010ai}. The non-trival thing is that these models are reproduced by a {\sl minimal} BI
model, i.e.~one which has only terms linear in the Einstein tensor in the determinant.
An expansion up to terms of mass dimension 8 has also been considered in \cite{Gullu:2010st}. To be explicit, the expansion of the Born-Infeld gravity theory \eqref{MBIM} including the mass dimension eight terms is given by
\bea
e^{-1} \cL^{(8)}_{{\text{BI}}} &=& - R + \ft1{m^2} \Big( R_{\m\n} R^{\m\n} - \ft38 R^2 \Big) + \ft{2}{3 m^4} \Big( R_{\m\n} R^{\n\r} R_\r{}^{\m} - \ft98 R R_{\m\n} R^{\m\n} + \ft{17}{64} R^3 \Big) \nn\\
&& - \ft{1}{8 m^6} \Big[R_{\m\n} R^{\n\r} R_{\r\s} R^{\s\m} - \ft53 R R_{\m\n}R^{\n\r} R_{\r}{}^\m + \ft{19}{16} R^2 R_{\m\n} R^{\m\n} \nn\\
&& - \ft14 (R_{\m\n} R^{\m\n})^2 - \ft{169}{768} R^4 \Big] \,.
\eea

Comparing the above  expansion of the Born-Infeld gravity theory with the results of this work,
it is interesting to speculate about the bosonic part of the $\cN=1$ Born-Infeld supergravity theory, in particular
about its dependence on the auxiliary scalar. Requiring supersymmetry  in lowest orders of a perturbative expansion leads us to consider
the following expression (without fermions and no cosmological constant)
\bea
I_{\text{SBI}}(G_{\m\n},S)  &=& +\,4 m^2 \int d^3 x\, \sqrt{-g} \Big( 1
+ \frac{2}{m^2} S^2 \Big) -\nonumber\\[.2truecm]
&&-\,4 m^2 \int d^3 x\,  \sqrt{-{\det}\Big(g_{\m\n} - \ft{1}{m^2} \big(G_{\m\n}-g_{\m\n}S^2\big) \Big)}  \,.
\label{SDBI0}
\eea
The perturbative expansion of this expression up to terms of mass dimension 8 is given by
\bea\label{above}
e^{-1} \cL^{(8)}_{\text{SBI}} &=& - (R + 2 S^2) + \ft{1}{m^2} \Big( R_{\m\n} R^{\m\n} - \ft38 R^2 - \ft12 R S^2 - \ft32 S^4 \Big) \nn\\ [.1truecm]
&& + \ft{2}{3 m^4} \Big(R_{\m\n} R^{\n\r} R_\r{}^{\m} - \ft98 R R_{\m\n} R^{\m\n} + \ft{17}{64} R^3  - \ft34 S^2 R_{\m\n} R^{\m\n} \nn\\ [.1truecm]
&& \qquad + \ft3{16} R S^4 + \ft9{32} R^2 S^6 + \ft38 S^6 \Big) \nn\\ [.1truecm]
&& - \ft{1}{8 m^6} \Big(R_{\m\n} R^{\n\r} R_{\r\s} R^{\s\m} - \ft53 R R_{\m\n}R^{\n\r} R_{\r}{}^\m + \ft{19}{16} R^2 R_{\m\n} R^{\m\n} \nn\\ [.1truecm]
&& \qquad - \ft14 (R_{\m\n} R^{\m\n})^2 - \ft{169}{768} R^4 - 2 R_{\m\n} R^{\n\r} R_{\r}{}^\m S^2 + \ft94 R R_{\m\n} R^{\m\n} S^2 \nn\\ [.1truecm]
&& \qquad + \ft34 R_{\m\n} R^{\m\n} S^4 - \ft{17}{32} R^3 S^2 - \ft18 R S^6 - \ft9{32} R^2 S^4 - \ft3{16} S^6 \Big)\,.
\eea
The above Lagrangian has many off-diagonal terms. Without a cosmological constant these terms are harmless.
The cosmological extension of  \eqref{SDBI0} can easily be obtained by adding the off-shell cosmological constant (\ref{CosmologicalConstant}). Now the off-diagonal terms are un-acceptable and they need to be canceled by subtracting from the above expression a number of off-diagonal invariants. The subtraction of these off-diagonal invariants not
only cancel the off-diagonal terms, they also change the coefficient in front of the purely auxiliary terms.
We have verified that the above expression \eqref{SDBI0}, after adding a cosmological constant and canceling all off-diagonal terms for the terms up to mass dimension 6  precisely yields the
bosonic part of (i) 3D $\cN=1$ Einstein supergravity; (ii) supersymmetric NMG and (iii) supersymmetric cubic extended NMG. In each case the purely gravitational and auxiliary part of these theories is reproduced.
What happens if we include the terms of mass dimension 8 is not yet known at this point. However, upon counting the number of off-diagonal terms and the number of possible off-diagonal invariants, both of mass dimension 8, we find that we have  enough invariants to cancel all the off-diagonal terms in the above eight derivative model.

In view of the above we propose that the bosonic part of the $\cN=1$ BI supergravity model is given by $I^\prime_{\text{SBI}}(G_{\m\n},S,M)$
where $I_{\text{SBI}}(G_{\m\n},S,M)$ indicates the cosmological extension, with a new cosmological mass parameter $M$, of $I_{\text{SBI}}(G_{\m\n},S)$, given in eq.~\eqref{SDBI0},
and  where the prime indicates that after expanding the expression of $I_{\text{SBI}}(G_{\m\n},S,M)$ one cancels
all off-diagonal terms by adding a number of off-diagonal invariants to the expression for $I_{\text{SBI}}(G_{\m\n},S,M)$.

As it stands, the expression for our proposal $I^\prime_{\text{SBI}}(G_{\m\n},S,M)$
 is not well-defined to all orders in an expansion of $1/m$. The reason for this is that one first has to know the expressions for all off-diagonal supersymmetric invariants that need to be subtracted in order to obtain a ghost-free model around the supersymmetric $AdS_3$ vacua. As mentioned above  these supersymmetric invariants will contain purely auxiliary terms with powers of $S$ that will modify the coefficients of similar terms that arise from the perturbative expansion of the Lagrangian \eqref{SDBI0}. Therefore, without the knowledge of all off-diagonal invariants, we cannot predict the final coefficients of the purely auxiliary terms. We only verified our proposal up to terms of mass dimension 6. To improve on this one should presumably work with a manifestly supersymmetric formulation using superfields.

\section{Conclusions}

In this paper we constructed the most general $\cN=1$ higher-derivative supergravity action containing terms
with no more than 6 derivatives. To achieve this, we made extensive use of the superconformal tensor calculus
from which we deduced several useful  auxiliary formulae such as multiplication rules of multiplets and
maps between different multiplets.

To avoid ghosts at the linearized level, we avoided terms of the  form $R\Box R$ and $R_{\m\n}\Box R^{\m\n}$.
We can easily extend our analysis and construct supersymmetric invariants that contain such terms.
Models with such terms generically contain two massive gravitons one of which is a ghost. However, there are critical values
of the parameters in which one or two of the massive gravitons become massless thereby avoiding the massive ghost.
In particular, the tri-critical gravity model, in which both massive gravitons become massless, was studied in
\cite{Bergshoeff:2012ev}. The tri-critical model of \cite{Bergshoeff:2012ev} was limited in the sense that it did not include terms cubic in the curvatures. We find that in the same way that the absence of a $S\Box S$ kinetic term fixes the NMG combination of $R^2$  and $R_{\m\n}R^{\m\n}$ terms, the absence of a higher-derivative $S\Box^2 S$ term
fixes the combination of
$R\Box R$ and $R_{\m\n}\Box R^{\m\n}$ terms to the one corresponding to tri-critical gravity. However, we also find that it is not possible to
cancel all off-diagonal terms in the supersymmetric extension of tri-critical gravity while keeping supersymmetry.\,\footnote{This is presumably related to the fact that
most likely tri-critical gravity at the nonlinear level contains ghosts \cite{Apolo:2012vv}.}

This leads us to consider the most general model containing {\sl all} possible curvature terms, with and without explicit derivatives. A particularly interesting combination of terms of mass dimension 6 is given by
\bea
e^{-1} \cL_{\text{Tri}} 
 &=& R_{\m\n} \Box R^{\m\n} - \ft38 R \Box R - 3 R_{\m\n} R^{\n\r} R_{\r}{}^\m + \ft52 R R_{\m\n} R^{\m\n} - \ft12 R^3 \,.
\eea
 Recently, it has been shown \cite{Afshar:2014new} that this combination  is free from  the non-linear Boulware-Deser ghosts
 \cite{Boulware:1973my}.
 We find that there is no supersymmetric extension of this invariant where all off-diagonal terms have been canceled.
 This is not surprising given the fact  that this model, due to the $R\Box R$ and $R_{\m\n}\Box R^{\m\n}$ terms,  contains a ghostly massive graviton.

It is intriguing that there are now several criteria that all hint to the same higher-derivative cubic extended NMG gravity model if we restrict to terms with at most 6 derivatives and require the absence of ghosts. We are aware of the following ones:
\bigskip

\noindent (1)\ {\sl Supersymmetry}\ \ In this work we have shown that supersymmetry (and
requiring the absence of ghosts when linearized around an $\text{AdS}_3$ background)
leads to a unique answer for the purely gravitational terms, see eq.~\eqref{SuperCNMG},  of mass dimension 6
corresponding to  the cubic extended NMG model.
\bigskip

\noindent (2)\ {\sl Holographic c-theorem}\ \
The same combination of cubic curvature terms \eqref{SuperCNMG}, also follows from considerations on the holographic c-theorem \cite{Sinha:2010ai}.
 \bigskip

\noindent (3)\  {\sl Born-Infeld Gravity}\ \ The cubic extended NMG model follows from a truncation of Born-Infeld gravity
\cite{Gullu:2010pc}. Unlike the previous two cases, the truncation of Born-Infeld gravity fixes the relative coefficients between the Einstein-Hilbert, quadratic, and cubic invariants. Therefore, the holographic c-theorem and supersymmetry leads to a wider class of ghost-free invariants then Born-Infeld gravity. In other words, Born-Infeld gravity corresponds to a specific choice of relative coefficients.
\bigskip

\noindent (4)\  {\sl Absence of Boulware-Deser Ghosts}\ \ There is one more criterion. Recently, it has been shown
in \cite{Afshar:2014new}
that precisely the cubic extended NMG model
follows from a Chern-Simons like formulation in which it is relatively easy to show the absence of non-linear ghosts using the Hamiltonian formalism \cite{Hohm:2012vh,Bergshoeff:2014bia}.
\bigskip

It is a challenge to see whether the Born-Infeld gravity model mentioned in criterion (3) passes the tests (1), (2) and (4).
As far as criterion (1) is concerned, one could  repeat the analysis of this paper to one level higher. However, for an all order result presumably a better formulation of Born-Infeld supergravity  using superspace techniques is needed \cite{Gates:2001ff}.
Concerning criterion (4) it would be interesting to see whether the full Born-Infeld gravity model allows for a Chern-Simons like formulation, see also \cite{Afshar:2014new}.

\section*{Acknowledgements}
We wish to thank H. Afshar, G.Alkac, L. Basanisi and W. Merbis for useful discussions. One of us (M.O.) is funded by a grant from Groningen University and the KNAW.


\begin{thebibliography}{99}



\bibitem{Lopes Cardoso:1998wt}
  G.~Lopes Cardoso, B.~de Wit and T.~Mohaupt,
  ``Corrections to macroscopic supersymmetric black hole entropy,''  Phys.\ Lett.\ B {\bf 451} (1999) 309  [hep-th/9812082].  



\bibitem{Maldacena:1997re}
  J.~M.~Maldacena,
  ``The Large N limit of superconformal field theories and supergravity,''  Adv.\ Theor.\ Math.\ Phys.\  {\bf 2} (1998) 231  [hep-th/9711200].  


\bibitem{Myers:2010tj}
  R.~C.~Myers and A.~Sinha,
  ``Holographic c-theorems in arbitrary dimensions,''  JHEP {\bf 1101} (2011) 125  [arXiv:1011.5819 [hep-th]].  


\bibitem{Bergshoeff:2009hq}
  E.~A.~Bergshoeff, O.~Hohm and P.~K.~Townsend,
``Massive Gravity in Three Dimensions,''
  Phys.\ Rev.\ Lett.\  {\bf 102}, 201301 (2009)


\bibitem{Deser:1981wh}
S.~Deser, R.~Jackiw, and S.~Templeton, ``{Topologically Massive Gauge
  Theories},'' {\em Annals Phys.} {\bf 140} (1982)
372--411.


\bibitem{Gullu:2010pc}
  I.~Gullu, T.~C.~Sisman and B.~Tekin,
``Born-Infeld extension of new massive gravity,''
  Class.\ Quant.\ Grav.\  {\bf 27} (2010) 162001
  [arXiv:1003.3935 [hep-th]].


\bibitem{Deser:1998rj}
  S.~Deser and G.~W.~Gibbons,
  ``Born-Infeld-Einstein actions?,''
  Class.\ Quant.\ Grav.\  {\bf 15}, L35 (1998)






\bibitem{Sinha:2010ai}
  A.~Sinha,
 ``On the new massive gravity and AdS/CFT,''
  JHEP {\bf 1006}, 061 (2010)

\bibitem{Paulos:2010ke}
  M.~F.~Paulos,
``New massive gravity extended with an arbitrary number of curvature corrections,''
  Phys.\ Rev.\ D {\bf 82}, 084042 (2010)
  [arXiv:1005.1646 [hep-th]].

\bibitem{Andringa:2009yc}
  R.~Andringa, E.~A.~Bergshoeff, M.~de Roo, O.~Hohm, E.~Sezgin and P.~K.~Townsend,
 ``Massive 3D Supergravity,''
  Class.\ Quant.\ Grav.\  {\bf 27}, 025010 (2010)



\bibitem{Bergshoeff:2010mf}
  E.~A.~Bergshoeff, O.~Hohm, J.~Rosseel, E.~Sezgin and P.~K.~Townsend,
 ``More on Massive 3D Supergravity,''
  Class.\ Quant.\ Grav.\  {\bf 28}, 015002 (2011)





\bibitem{Nutma:2012ss}
  T.~Nutma,
  ``Polycritical Gravities,''  Phys.\ Rev.\ D {\bf 85} (2012) 124040  [arXiv:1203.5338 [hep-th]].  




\bibitem{Gates:1983nr}
  S.~J.~Gates, M.~T.~Grisaru, M.~Rocek and W.~Siegel,
  ``Superspace Or One Thousand and One Lessons in Supersymmetry,''  Front.\ Phys.\  {\bf 58} (1983) 1  [hep-th/0108200].  




\bibitem{vanNieuwenhuizen:1985cx}
  P.~van Nieuwenhuizen,
 ``$D=3$ Conformal Supergravity and Chern-simons Terms,''
  Phys.\ Rev.\ D {\bf 32}, 872 (1985).

\bibitem{Uematsu:1984zy}
  T.~Uematsu,
 ``Structure of $N=1$ Conformal and Poincare Supergravity in (1+1)-dimensions and (2+1)-dimensions,''
  Z.\ Phys.\ C {\bf 29}, 143 (1985).

\bibitem{Uematsu:1986de}
  T.~Uematsu,
  ``Constraints and Actions in Two-dimensional and Three-dimensional $N=1$ Conformal Supergravity,''
  Z.\ Phys.\ C {\bf 32}, 33 (1986).

\bibitem{Howe:1977us}
  P.~S.~Howe and R.~W.~Tucker,
 ``Local Supersymmetry in (2+1)-Dimensions. 1. Supergravity and Differential Forms,''
  J.\ Math.\ Phys.\  {\bf 19}, 869 (1978).

\bibitem{Gullu:2010st}
  I.~Gullu, T.~C.~Sisman and B.~Tekin,
``c-functions in the Born-Infeld extended New Massive Gravity,''
  Phys.\ Rev.\ D {\bf 82}, 024032 (2010)


\bibitem{Bergshoeff:2012ev}
  E.~A.~Bergshoeff, S.~de Haan, W.~Merbis, J.~Rosseel and T.~Zojer,
``On Three-Dimensional Tricritical Gravity,''
  Phys.\ Rev.\ D {\bf 86}, 064037 (2012)

\bibitem{Apolo:2012vv}
  L.~Apolo and M.~Porrati,
  ``Nonlinear Dynamics of Parity-Even Tricritical Gravity in Three and Four Dimensions,''  JHEP {\bf 1208} (2012) 051  [arXiv:1206.5231 [hep-th]].  




\bibitem{Afshar:2014new}
  H.~R.~Afshar, E.~A.~Bergshoeff and W.~Merbis,
  ``Extended massive gravity in three dimensions,''
  arXiv:1405.6213 [hep-th].


\bibitem{Boulware:1973my}
  D.~G.~Boulware and S.~Deser,
  ``Can gravitation have a finite range?,''  Phys.\ Rev.\ D {\bf 6} (1972) 3368.  


\bibitem{Hohm:2012vh}
  O.~Hohm, A.~Routh, P.~K.~Townsend and B.~Zhang,
  ``On the Hamiltonian form of 3D massive gravity,''  Phys.\ Rev.\ D {\bf 86} (2012) 084035  [arXiv:1208.0038].  

\bibitem{Bergshoeff:2014bia}
  E.~A.~Bergshoeff, O.~Hohm, W.~Merbis, A.~J.~Routh and P.~K.~Townsend,
  ``The Hamiltonian Form of Three-Dimensional Chern-Simons-like Gravity Models,''  arXiv:1402.1688 [hep-th].  

\bibitem{Gates:2001ff} 
  S.~J.~Gates, Jr. and S.~V.~Ketov,
  ``4-D, N=1 Born-Infeld supergravity,''
  Class.\ Quant.\ Grav.\  {\bf 18}, 3561 (2001)


\end{thebibliography}
\end{document}